 \documentclass[preprint,superscriptaddress,showpacs,nofootinbib,aps]{revtex4}
 \usepackage{graphicx}
 \usepackage{bm}
 \begin{document}

 \title{Phenomenological Analysis of Charmless\\
        Decays $B_{s}{\to}PP,PV$ with QCD Factorization}
 \thanks{Supported in part by National Natural Science Foundation
         of China}
 \author{Junfeng Sun}
 \email[E-mail address: ]{sunjf@mail.ihep.ac.cn}
 \affiliation{Institute of High Energy Physics,
              Chinese Academy of Sciences,\\
              P.O.Box 918(4),
              Beijing 100039, China}
 \author{Guohuai Zhu}
 \email[E-mail address: ]{zhugh@post.kek.jp.}
 \affiliation{Theory Group, KEK, Tsukuba,
              Ibaraki 305-0801, Japan}
 \author{Dongsheng Du}
 \email[E-mail address: ]{duds@mail.ihep.ac.cn}
 \affiliation{Institute of High Energy Physics,
              Chinese Academy of Sciences,\\
              P.O.Box 918(4),
              Beijing 100039, China}
 \date{\today}

 \begin{abstract}
 We calculated the $CP$ averaged branching ratios and $CP$-violating
 asymmetries of two-body charmless hadronic $B_{s}$ ${\to}$ $PP$, $PV$
 decays with the QCDF approach, including the contributions from the
 chirally enhanced power  corrections  and  weak annihilations. Only
 several decay modes, such as
 $B_{s}$ ${\to}$ $K^{({\ast})}K$, $K^{({\ast}){\pm}}{\pi}^{\mp}$,
 $K^{\pm}{\rho}^{\mp}$, ${\eta}^{(\prime)}{\eta}^{(\prime)}$,
 have large branching ratios, which may be observed in the near future.
 The penguin-to-tree ratio
 ${\vert}P_{{\pi}{\pi}}/T_{{\pi}{\pi}}{\vert}$ and a bound on angle
 ${\gamma}$ are given.
 \end{abstract}
 \pacs{13.25.Hw 12.38.Bx}

 \maketitle

 \section{Introduction}
 \label{sec1}
 Recently there has been remarkable progress in the study of exclusive
 charmless $B_{u,d}$ decays. Experimentally, many two-body non-leptonic
 charmless $B_{u,d}$ decays have been observed by CLEO and $B$-factories
 at KEK and SLAC (see Refs. \cite{0111041,0202022,0206053,0207055,%
 0207065,0111037,0204002,0207007,0207033,0207090}),
 and more $B$ decay channels will be measured with great precision soon.
 With the accumulation of data, the Standard Model can be tested in more
 detail. Theoretically, several attractive methods have been proposed
 to study the nonfactorizable effects in hadronic matrix elements from
 first principles, such as QCD factorization (QCDF) \cite{9905312},
 perturbative QCD method (PQCD) \cite{9607214,9701233,0004004}, and so on.
 Intensive investigations on hadronic charmless two-body $B_{u,d}$ decays
 have been studied in detail, for example, in Refs.
 \cite{9804363,9903453,0012208,0005006,0104090,0104110,0108141,0201253}.

 The potential $B_{s}$ decay modes permit us to overconstrain the
 unitarity CKM matrix. This makes the search for $CP$ violation in the
 $B_{s}$ decays highly interesting. The problem is that $B_{s}$ mesons
 oscillate at a high frequency, and nonleptonic $B_{s}$ decays have still
 remained elusive from observation. Today only some weak upper limits on
 branching ratios of several charmless hadronic decays are available,
 mostly from LEP and SLD experiments \cite{pdg2002}, such as $B_{s}^{0}$
 ${\to}$ ${\pi}^{+}{\pi}^{-}$, ${\pi}^{0}{\pi}^{0}$, ${\eta}{\pi}^{0}$,
 ${\eta}{\eta}$, $K^{+}K^{-}$, ${\pi}^{+}K^{-}$, ${\cdots}$. Unlike
 $B_{u,d}$ mesons, the heavier $B_{s}$ mesons cannot be studied at the
 $B$-factories operating at the ${\Upsilon}(4S)$ resonance. However it
 is believed that in the future at hadron colliders, such as CDF, D0,
 HERA-B, BTeV, and LHCb, the signs of $CP$ violation in $B_{s}$ system
 can be observed with high accuracy in addition to studies of certain
 $B_{u,d}$ modes \cite{0201071}.

 The early theoretical studies of two-body charmless nonleptonic decays of
 $B_{s}$ meson can be found in Refs. \cite{prd38,plb318,plb320,prd48,9503278}.
 The investigation on the exclusive charmless $B_{s}$ decays into final
 states containing ${\eta}^{(\prime)}$ meson was given within the
 generalized factorization framework \cite{9807393}. Chen, Cheng, and
 Tseng calculated carefully the branching ratios for charmless decays
 $B_{s}$ ${\to}$ $PP$, $PV$, $VV$ (here $P$ and $V$ denote pseudoscalar
 and vector mesons respectively) \cite{9809364}. And new physics effects
 in $B_{s}$ decays was considered in \cite{0012063}. It is found that the
 electroweak penguin contributions can be large for some decays modes
 \cite{9503278,9809364}, and that branching ratios for $B_{s}$ ${\to}$
 ${\eta}{\eta}^{\prime}$ and several other decay modes can be as large as
 $10^{-5}$ \cite{9807393,9809364,0012063} which is measurable at future
 experiments.

 Few years ago, Beneke, Buchalla, Neubert, and Sachrajda gave a QCDF
 formula to compute the hadronic matrix elements
 ${\langle}M_{1}M_{2}{\vert}O_{i}{\vert}B{\rangle}$
 in the heavy quark limit, so that the hadronic uncertainties enter only
 at the level of power corrections of $1/m_{b}$. This basic
 formula is presumed to be valid for $B$ decays into two light final
 states \cite{9905312,0102077}. We made a comprehensive analysis on
 exclusive hadronic $B_{u,d}$ decay using the QCDF approach, and
 calculated the branching ratios and CP asymmetries for decays $B_{u,d}$
 ${\to}$ $PP$ \cite{0108141} and $PV$ \cite{0201253}. We find that with
 appropriate parameters, most of our predictions are in agreement with the
 present experimental data. In this paper, we would like to apply the QCDF
 approach to the case of $B_{s}$ mesons.

 This paper is organized as follow: In section \ref{sec2}, we discuss
 the theoretical framework and define the relevant matrix elements for
 $B_{s}$ ${\to}$ $PP$, $PV$ decays. In section \ref{sec3}, we list the
 theoretical input parameters used in our analysis. Section \ref{sec4}
 and section \ref{sec5} are devoted to the numerical results and some
 remarks of $CP$ averaged branching ratios and $CP$-violating asymmetries,
 respectively. In the mean time, the theoretical uncertainties due to
 the variation of inputs are investigated. In section \ref{sec6}, we
 give the values of the penguin-to-tree ratio
 $P_{{\pi}{\pi}}/T_{{\pi}{\pi}}$ and a constraint on weak angle
 ${\gamma}$. Finally, we conclude with a summary in section \ref{sec7}.

 \section{Theoretical framework for B decays}
 \label{sec2}

 \subsection{The effective Hamiltonian}
 \label{sec21}
 Using the operator product expansion and renormalization group equation,
 the low energy effective Hamiltonian relevant to nonleptonic $B$ decays
 can be written as \cite{9512380}:
 \begin{eqnarray}
 {\cal H}_{eff} &=& \frac{G_{F}}{\sqrt{2}}
   {\sum\limits_{q=u,c}} v_{q} \Big\{
     C_{1}({\mu}) Q^{q}_{1}({\mu})
  +  C_{2}({\mu}) Q^{q}_{2}({\mu})
  +{\sum\limits_{k=3}^{10}} C_{k}({\mu}) Q_{k}({\mu})
    \nonumber \\
 &+& C_{7{\gamma}} Q_{7{\gamma}}
  +  C_{8g} Q_{8g}  \Big\} + \text{H.c.} ,
 \label{eq:Hamiltonian}
 \end{eqnarray}
 where $v_{q}$ $=$ $V_{qb}$ $V_{qd}^{\ast}$ (for $b{\to}d$ transition) or
 $v_{q}$ $=$ $V_{qb}$ $V_{qs}^{\ast}$ (for $b{\to}s$ transition) are CKM
 factors. $C_{i}({\mu})$ are Wilson coefficients which have been reliably
 evaluated to the next-to-leading logarithmic order. Their numerical
 values in the naive dimensional regularization scheme at three different
 scales are listed in Table \ref{tab1}. The effective operators, $Q_{i}$,
 can be expressed explicitly as follows:
 \begin{eqnarray}
  & &Q^{u}_{1}=({\bar{u}}_{\alpha}b_{\alpha})_{V-A}
               ({\bar{q}}_{\beta} u_{\beta} )_{V-A},
     \ \ \ \ \ \ \ \ \ \ \ \ \ \ \ \ \ \ \
     Q^{c}_{1}=({\bar{c}}_{\alpha}b_{\alpha})_{V-A}
               ({\bar{q}}_{\beta} c_{\beta} )_{V-A},
  \label{eq:operator-q1} \\
  & &Q^{u}_{2}=({\bar{u}}_{\alpha}b_{\beta} )_{V-A}
               ({\bar{q}}_{\beta} u_{\alpha})_{V-A},
     \ \ \ \ \ \ \ \ \ \ \ \ \ \ \ \ \ \ \
     Q^{c}_{2}=({\bar{c}}_{\alpha}b_{\beta} )_{V-A}
            ({\bar{q}}_{\beta} c_{\alpha})_{V-A},
  \label{eq:operator-q2} \\
  & &Q_{3}=({\bar{q}}_{\alpha}b_{\alpha})_{V-A}\sum\limits_{q^{\prime}}
           ({\bar{q}}^{\prime}_{\beta} q^{\prime}_{\beta} )_{V-A},
     \ \ \ \ \ \ \ \ \ \ \ \ \ \ \ \
     Q_{4}=({\bar{q}}_{\beta} b_{\alpha})_{V-A}\sum\limits_{q^{\prime}}
           ({\bar{q}}^{\prime}_{\alpha}q^{\prime}_{\beta} )_{V-A},
  \label{eq:operator-q3-q4} \\
  & &Q_{5}=({\bar{q}}_{\alpha}b_{\alpha})_{V-A}\sum\limits_{q^{\prime}}
           ({\bar{q}}^{\prime}_{\beta} q^{\prime}_{\beta} )_{V+A},
     \ \ \ \ \ \ \ \ \ \ \ \ \ \ \ \
     Q_{6}=({\bar{q}}_{\beta} b_{\alpha})_{V-A}\sum\limits_{q^{\prime}}
           ({\bar{q}}^{\prime}_{\alpha}q^{\prime}_{\beta} )_{V+A},
  \label{eq:operator-q5-q6} \\
  & &Q_{7}=\frac{3}{2}({\bar{q}}_{\alpha}b_{\alpha})_{V-A}
           \sum\limits_{q^{\prime}}e_{q^{\prime}}
           ({\bar{q}}^{\prime}_{\beta} q^{\prime}_{\beta} )_{V+A},
     \ \ \ \ \ \ \ \ \ \
     Q_{8}=\frac{3}{2}({\bar{q}}_{\beta} b_{\alpha})_{V-A}
           \sum\limits_{q^{\prime}}e_{q^{\prime}}
           ({\bar{q}}^{\prime}_{\alpha}q^{\prime}_{\beta} )_{V+A},
  \label{eq:operator-q7-q8} \\
  & &Q_{9}=\frac{3}{2}({\bar{q}}_{\alpha}b_{\alpha})_{V-A}
           \sum\limits_{q^{\prime}}e_{q^{\prime}}
           ({\bar{q}}^{\prime}_{\beta} q^{\prime}_{\beta} )_{V-A},
     \ \ \ \ \ \ \ \ \ \
    Q_{10}=\frac{3}{2}({\bar{q}}_{\beta} b_{\alpha})_{V-A}
           \sum\limits_{q^{\prime}}e_{q^{\prime}}
           ({\bar{q}}^{\prime}_{\alpha}q^{\prime}_{\beta} )_{V-A},
  \label{eq:operator-q9-q10} \\
  & &Q_{7{\gamma}}=\frac{e}{8{\pi}^{2}}m_{b}{\bar{q}}_{\alpha}
           {\sigma}^{{\mu}{\nu}}(1+{\gamma}_{5})
            b_{\alpha}F_{{\mu}{\nu}},
     \ \ \ \ \ \ \ \ \ \ \
     Q_{8g}=\frac{g}{8{\pi}^{2}}m_{b}{\bar{q}}_{\alpha}
           {\sigma}^{{\mu}{\nu}}(1+{\gamma}_{5})
            t^{a}_{{\alpha}{\beta}}b_{\beta}G^{a}_{{\mu}{\nu}},
 \label{eq:operator-q7r-q8g}
 \end{eqnarray}
 where $q^{\prime}$ denotes all the active quarks at scale ${\mu}$ $=$
 ${\cal O}(m_{b})$, i.e. $q^{\prime}$ $=$ $u$, $d$, $s$, $c$, $b$.

 \subsection{Hadronic matrix elements within the QCDF framework}
 \label{sec22}
 To get the decay amplitudes, the most difficult theoretical work
 is to compute  the  hadronic matrix elements of the effective
 operators, i.e. ${\langle}M_{1}M_{2}{\vert}O_{i}{\vert}B{\rangle}$.
 Phenomenologically, these hadronic matrix elements are usually
 parameterized into the product of the decay constants and the
 transition form factors based on the naive factorization scheme
 (NF) \cite{bsw}. However, one main defect of the rough NF approach
 is that hadronic matrix elements cannot make compensation for the
 renormalization scheme- and scale- dependence of Wilson coefficients,
 in this sense NF's results are unphysical. This indicates that
 ``nonfactorizable'' contributions from high order corrections to
 the hadronic matrix elements must be taken into account.

 The QCDF approach is one of novel methods to evaluate these hadronic
 matrix elements relevant to $B$ decays systematically. In the heavy
 quark limit $m_{b}$ $\gg$ ${\Lambda}_{QCD}$, up to power corrections
 of order of ${\Lambda}_{QCD}/m_{b}$, the basic QCDF formula is
 \cite{9905312}
 \begin{eqnarray}
 {\langle}M_{1}M_{2}{\vert}O_{i}{\vert}B{\rangle} &=&
   \sum\limits_{j} F^{B{\to}M_{1}}_{j} {\int}_{0}^{1}dx \
   T^{I}_{ij}(x) {\Phi}_{M_{2}}(x) + ( M_{1} {\leftrightarrow} M_{2} )
   \nonumber \\ & & \!\!\!\! \!\!\!\! \!\!\!\! \!\!\!\!
 + {\int}_{0}^{1} d{\xi} {\int}_{0}^{1} dx {\int}_{0}^{1} dy \
   T^{II}_{i}({\xi},x,y) {\Phi}_{B}(\xi) {\Phi}_{M_{1}}(x)
   {\Phi}_{M_{2}}(y) \nonumber \\ &=&
 {\langle}M_{1}M_{2}{\vert}J_{1}{\otimes}J_{2}{\vert}B{\rangle}_{F}
  \Big[ 1 + {\sum} r_{n} {\alpha}_{s}^{n}
          + {\cal O}({\Lambda}_{QCD}/m_{b}) \Big]
 \label{eq:qcdf}
 \end{eqnarray}
 where $T^{I,II}_{i}$ denote hard-scattering kernels. At leading order,
 $T^{I}_{i}=1$, $T^{II}_{i}=0$, the QCDF formula (\ref{eq:qcdf}) shows
 that there is no long-distance interaction between $M_{2}$ meson and
 ($BM_{1}$) system, and reproduces the NF's results. Neglecting the power
 corrections of ${\cal O}({\Lambda}_{QCD}/m_{b})$, $T^{I,II}_{i}$ are
 hard gluon exchange dominant, and therefore calculable order by order
 with perturbative theory. Nonperturbative effects are either
 suppressed by $1/m_{b}$ or parameterized in terms of mesons decay
 constants, form factors $F^{B{\to}M}$, and meson light-cone distribution
 amplitudes ${\Phi}_{B}(\xi)$, ${\Phi}_{M}(x)$. The factorized matrix
 elements
 ${\langle}M_{1}M_{2}{\vert}J_{1}{\otimes}J_{2}{\vert}B{\rangle}_{F}$
 is the same as the definition of the BSW approximation \cite{bsw}.
 Through the QCDF formula, the hadronic matrix elements can be separated
 into short-distance part and long-distance part, and the ``residual''
 renormalization scheme- and scale- dependence of hadronic matrix
 elements could be extracted to cancel those of the corresponding Wilson
 coefficients, so that physical results at least at the order of
 ${\alpha}_{s}$ level are renormalization scheme and scale independent
 \cite{0102077}. Through the QCDF formula, ``nonfactorizable'' effects
 can be evaluated, and partial information about the strong phases can
 be obtained.

 It is important to note that some power suppression might fail in some
 cases because the $b$ quark mass is not asymptotically large. For
 example, power correction proportional to $2m_{M}^2/(m_{b}m_{q})$
 with $q=$ $u$, $d$, $s$, which is formally power suppressed, is now
 chirally enhanced and numerically important to penguin dominated $B$
 rare decays. Therefore it is necessary to include at least the chirally
 enhanced corrections consistently for phenomenological application of
 QCDF in $B$ decays. However, the twist-3 corrections
 to hard scattering kernels $T^{II}$ cannot provide sufficient
 endpoint suppression, so there appears infrared logarithmic divergence,
 ${\int}dx/x$ ${\sim}$ ${\ln}(m_{b}/{\Lambda}_{QCD})$. In PQCD method,
 this singularity can be smoothed out by introducing the partonic
 intrinsic transverse momentum and the mechanism of Sudakov suppression.
 It should be interesting to investigate the possibility of incorporating the
 Sudakov form factor into the QCDF approach \cite{0107320}. But it is really
 uneasy because the Sudakov suppression is one of the key ideas of the PQCD
 method while PQCD and QCDF have different power expansions which lead to
 completely different understanding on $B$ decays. Therefore, to take the
 chirally enhanced corrections into account,
 in this paper we adopt phenomenological treatment for the divergent integral
 ${\int}dx/x$ \cite{0104110}.

 For weak annihilation contributions, they are believed to be very small with
 the naive factorization assumption (see, for example Ref. \cite{9804363}).
 Within the QCDF approach, the weak annihilation amplitudes are also formerly
 suppressed by $(f_{B}f_{M_{1}})/(F^{B{\to}M_{1}}m_{B}^{2})$ ${\sim}$
 $\Lambda_{QCD}/m_b$ \cite{9905312}. But as emphasized in the PQCD method
 \cite{0004004,0006001,0004213},
 annihilation contributions with QCD corrections could give potentially large
 strong phases, hence large $CP$ violation could be expected. In addition, the
 phenomenological investigations on $B$ decays \cite{0104110,0108141,0201253}
 within the QCDF framework also suggest that their effects could be sizable
 when large model uncertainties are considered. So annihilation contributions
 cannot be simply neglected.
 In the previous work \cite{0104110,0108141,0201253} it has been shown that
 annihilation contributions exhibit endpoint singularities even with
 leading twist distribution amplitudes for the final states, and these
 infrared divergence must be parameterized, so extra theoretical
 uncertainties and model dependence are introduced. In spite of these
 problems, it is still interesting to
 estimate the weak annihilation effects in this paper.

 With the QCDF approach, matrix elements for two-body $B_{s}$ decays
 can be written as,
 \begin{eqnarray}
 {\langle}M_{1}M_{2}{\vert} {\cal H}_{eff} {\vert}B_{s}{\rangle} &=&
 {\cal A}^{f}(B_{s}{\to}M_{1}M_{2}) + {\cal A}^{a}(B_{s}{\to}M_{1}M_{2}),
 \label{eq:am-01} \\
 {\cal A}^{f}(B_{s}{\to}M_{1}M_{2}) &=& \frac{G_{F}}{\sqrt{2}}
    \sum\limits_{q=u,c} \sum\limits_{i=1}^{10} v_{q} a_{i}^{q}
   {\langle}M_{1}M_{2}{\vert}J_{1}{\otimes}J_{2}{\vert}B_{s}{\rangle}_{F},
 \label{eq:am-02} \\
 {\cal A}^{a}(B_{s}{\to}M_{1}M_{2}) &{\propto}& \frac{G_{F}}{\sqrt{2}}
    \sum\limits_{q=u,c} \sum\limits_{i}
    f_{B_{s}} f_{M_{1}} f_{M_{2}} v_{q} b_{i},
 \label{eq:am-03}
 \end{eqnarray}
 where ${\cal A}^{a}$ arises from weak annihilation contributions.
 $f_{B_{s}}$, $f_{M}$ are decay constants for $B_{s}$, and $M$ mesons,
 respectively. ($M$ can be either pseudoscalar meson $P$ or vector meson
 $V$) The explicit expressions of decay amplitudes ${\cal A}^{f,a}$
 for $B_{s}$ ${\to}$ $PP$, $PV$ are listed in Appendix \ref{sec:app1},
 \ref{sec:app2}, \ref{sec:app3}, \ref{sec:app4}. Summary of the dynamical
 quantities $a_{i}$, $b_{i}$ is given in Refs. \cite{0104110,0108141,%
 0201253}.

 \section{Input parameters}
 \label{sec3}
 Theoretical expressions for decay amplitudes using the QCDF
 approach are complicated and depend on many input parameters including
 the SM parameters (such as CKM matrix elements, quark masses), Wilson
 coefficients and the renormalization scale ${\mu}$, and some soft and
 nonperturbative hadronic quantities (such as meson decay constants, form
 factors, and meson light cone distribution amplitudes), and so on. If
 quantitative predictions are to be made, the values for various
 parameters employed in this paper must be specified. It has been shown
 that the renormalization scale dependence has been greatly reduced
 compared to the NF's coefficients $a_{i,I}$ obtained at leading order
 \cite{0104110,0108141,0201253}, and the residual scale dependence should
 be further reduced when the higher order radiative corrections are
 considered. In calculations we use ${\mu}=m_{b}$. The rest parameters are
 discussed below.

 \subsection{The CKM matrix elements}
 \label{sec31}
 One widely used approximate form of the CKM matrix is the Wolfenstein
 parameterization \cite{ckm} which emphasizes the hierarchy among its
 elements and expresses them in terms of powers of
 ${\lambda}=$ ${\vert}V_{us}{\vert}$,
 \begin{equation}
 V_{CKM}=\left( \begin{array}{ccc}
     1-{\lambda}^{2}/2
  &    {\lambda}
  &   A{\lambda}^{3}({\rho}-i{\eta})  \\
      -{\lambda}
  &  1-{\lambda}^{2}/2
  &   A{\lambda}^{2}                  \\
      A{\lambda}^{3}(1-{\rho}-i{\eta})
  &  -A{\lambda}^{2}
  &  1  \end{array} \right) + {\cal O}({\lambda}^{4}).
 \label{eq:ckm}
 \end{equation}
 The values of four Wolfenstein parameters ($A$, ${\lambda}$, ${\rho}$,
 and ${\eta}$) are given by several analysis methods from the best
 knowledge of the experimental and theoretical inputs (for example,
 see Table \ref{tab2}). Within one standard deviation, the results
 from different approaches \cite{pdg2002,0012308,0104062,0207101} are
 virtually consistent with each other. In this paper, we shall take
 ${\lambda}$ = 0.2236 ${\pm}$ 0.0031,
 $A$         = 0.824  ${\pm}$ 0.046,
 $\bar{\rho}$= 0.22   ${\pm}$ 0.10,
 $\bar{\eta}$= 0.35   ${\pm}$ 0.05, and
 ${\gamma}$  = $(59{\pm}13)^{\circ}$ \cite{pdg2002}.

 However, it is not the right time to draw definite conclusions on the
 parameters ${\rho}$, ${\eta}$ and ${\gamma}$. Some interesting hints
 seem to favour ${\gamma}{\gtrsim}90^{\circ}$, which is in conflict
 with the data in Table \ref{tab2}. For example, it is assumed that it
 is possible to derive constraint on the angle ${\gamma}$ from a global
 analysis $B$ decays. Bargiotti, {\em et al.} obtain the bound
 ${\vert}{\gamma} - 90^{\circ}{\vert} > 21^{\circ}$ at $95\%$ C.L.
 from $B$ ${\to}$ $K{\pi}$ decay rates and $CP$ asymmetries \cite{0204029}.
 Combining the QCDF approach with a global CKM matrix analysis --- Rfit
 scenario advocated in \cite{0104062}, Beneke and Neubert make a fit of
 the Wolfenstein parameters $({\bar{\rho}},{\bar{\eta}})$ to six $B$
 ${\to}$ ${\pi}{\pi}$, $K{\pi}$ decays with updated measurements, and
 their fit tend to favor ${\gamma} > 90^{\circ}$ \cite{0207228}.
 In analogy with works \cite{0207228}, we also make a fit of the CKM
 matrix parameters to $B$ ${\to}$ $PP$, $PV$ decays, and the preliminary
 results are ${\lambda}$ = 0.22, $A$ = 0.82, ${\bar{\rho}}$ = 0.086,
 ${\bar{\eta}}$ = 0.39, and ${\gamma}$ = $78.8^{\circ}$ \cite{0209233}.
 For comparison, we shall take the results in \cite{0209233} as the CKM
 matrix inputs.

 \subsection{Quark masses}
 \label{sec32}
 There are two different classes of quark masses. One type is pole mass
 for constituent quark, which appears in the penguin loop corrections
 with the functions $G_{M}(s_{q})$ and ${\hat{G}}_{M}(s_{q})$, where
 $s_{q}$ = $m_{q}^{2}/m_{b}^{2}$. The definitions of $G_{M}(s_{q})$ and
 ${\hat{G}}_{M}(s_{q})$ can be found in \cite{0104110}. In this paper,
 we take
 \begin{equation}
  m_{u}=m_{d}=m_{s}=0,   \ \ \ \
  m_{c}=1.47 \; \text{GeV}, \ \ \ \
  m_{b}=4.66 \; \text{GeV}.
 \label{eq:mass-01}
 \end{equation}
 The other is current quark mass which appears in the equations of
 motions, and is renormalization scale dependent. Their values are
 \cite{pdg2002}
 \begin{eqnarray}
 & & \frac{1}{2} \Big[ {\overline{m}}_{u}(2\,\text{GeV})
                     + {\overline{m}}_{d}(2\,\text{GeV}) \Big]
        = (4.2{\pm}1.0)\,\text{MeV}, \label{eq:mass-q} \\
 & & {\overline{m}}_{s}(2\,\text{GeV})
        = (105{\pm}25)\,\text{MeV},  \label{eq:mass-s} \\
 & & {\overline{m}}_{b}({\overline{m}}_{b})
        = (4.26{\pm}0.15{\pm}0.15)\,\text{GeV}. \label{eq:mass-b}
 \end{eqnarray}
 Here we would like to use their central values for discussion. And using
 the renormalization group equation, their corresponding values at the
 scale of ${\mu}={\cal O}(m_{b})$ can be obtained. Because the current
 masses of light quarks are determined with large uncertainties, for
 illustration, we take approximation
 \begin{equation}
 r_{\chi}       = \frac{2{\mu}_{P}}{ {\overline{m}}_{b} } =
 r_{\chi}^{\pi} =
 r_{\chi}^{{\eta}^{({\prime})}} \Big( 1-
    \frac{f_{{\eta}^{({\prime})}}^{u}}{f_{{\eta}^{({\prime})}}^{s}} \Big)=
 r_{\chi}^{K}   =
    \frac{2m_{K}^{2}}{ {\overline{m}}_{b}
 ( {\overline{m}}_{s} + {\overline{m}}_{q} ) },
 \label{eq:mass-02}
 \end{equation}

 \subsection{Nonperturbative hadronic quantities}
 \label{sec33}
 Nonperturbative hadronic quantities, such as meson decay constants, form
 factors, and meson light cone distribution amplitudes, appear as inputs
 in the QCDF formula (\ref{eq:qcdf}). In principle, information about decay
 constants and form factors can be determined form experiments and/or
 theoretical estimations. Now we specify these parameters. In this paper,
 we assume ideal mixing between $\omega$ and $\phi$, i.e.
 ${\omega}$ = $(u\bar{u}+d\bar{d})/{\sqrt{2}}$ and ${\phi}$ = $s\bar{s}$.
 As to ${\eta}$ and ${\eta}^{\prime}$, we take the convention in
 \cite{9804363,9802409}, using two-mixing-angle formula for the decay
 constants, but without the charm quark content in ${\eta}$ and
 ${\eta}^{\prime}$.
 \begin{eqnarray}
 & &{\langle}0{\vert}\bar{q}{\gamma}_{\mu}{\gamma}_{5}q
           {\vert}{\eta}^{(\prime)}(p){\rangle}
          =if^{q}_{{\eta}^{(\prime)}}p_{\mu},
    \ \ \ \ \ \ \ (q=u,d,s)
 \label{eq:eta-01} \\
 & &\frac{{\langle}0{\vert}\bar{u}{\gamma}_{5}u
                    {\vert}{\eta}^{(\prime)}{\rangle}}
         {{\langle}0{\vert}\bar{s}{\gamma}_{5}s
                    {\vert}{\eta}^{(\prime)}{\rangle}}
  = \frac{f^{u}_{{\eta}^{(\prime)}}}{f^{s}_{{\eta}^{(\prime)}}},
    \ \ \ \ \ \
          {\langle}0{\vert}\bar{s}{\gamma}_{5}s
                    {\vert}{\eta}^{(\prime)}{\rangle}
  = -i\frac{m_{{\eta}^{(\prime)}}^{2}}{2m_{s}}
   (f^{s}_{{\eta}^{(\prime)}}-f^{u}_{{\eta}^{(\prime)}}),
 \label{eq:eta-02} \\
 & & f^{u}_{{\eta}^{\prime}}
     = \frac{ f_{8} }{ \sqrt{6} } {\sin}{\theta}_{8}
     + \frac{ f_{0} }{ \sqrt{3} } {\cos}{\theta}_{0},
   \ \ \ \ \ \ \ \
     f^{s}_{{\eta}^{\prime}}
     = -2 \frac{ f_{8} }{ \sqrt{6} } {\sin}{\theta}_{8}
     +    \frac{ f_{0} }{ \sqrt{3} } {\cos}{\theta}_{0},
 \label{eq:eta-03} \\
 & & f^{u}_{\eta} =  \frac{ f_{8} }{ \sqrt{6} } {\cos}{\theta}_{8}
              -  \frac{ f_{0} }{ \sqrt{3} } {\sin}{\theta}_{0},
   \ \ \ \ \ \ \ \
     f^{s}_{\eta} = -2 \frac{ f_{8} }{ \sqrt{6} } {\cos}{\theta}_{8}
                -  \frac{ f_{0} }{ \sqrt{3} } {\sin}{\theta}_{0},
 \label{eq:eta-04}
 \end{eqnarray}
 And for $B_{s}$ ${\to}$ ${\eta}^{({\prime})}$ transition form factors,
 we take \cite{9809364}
 \begin{equation}
 F_{0}^{B_{s}{\eta}}
     = - \Big( \frac{2}{\sqrt{6}} {\cos}{\theta}
             + \frac{1}{\sqrt{3}} {\sin}{\theta} \Big)
       F_{0}^{B_{s}{\eta}_{s\bar{s}}},
 \ \ \ \ \ \
  F_{0}^{B_{s}{\eta}^{\prime}}
     = \Big( - \frac{2}{\sqrt{6}} {\sin}{\theta}
             + \frac{1}{\sqrt{3}} {\cos}{\theta} \Big)
       F_{0}^{B_{s}{\eta}^{\prime}_{s\bar{s}}},
 \label{eq:eta-05}
 \end{equation}
 The values of these parameters are collected in Table \ref{tab3}.

 In this paper, we consider the contributions from chirally enhanced
 twist-3 light cone distribution amplitudes of a light pseudoscalar
 meson. As to vector mesons, only the longitudinally polarized twist-2
 terms are taken into account, and the effects from transversely polarized
 and higher twist parts are neglected because they are power suppressed.
 In calculation, we shall take their asymptotic forms, as displayed in
 \cite{0201253}, i.e. for a light pseudoscalar meson, we have
 \cite{0104110,0008255}:
 \begin{eqnarray}
 & &{\langle}P(k){\vert}\bar{q}(z_{2})q(z_{1}){\vert}0{\rangle}
    \nonumber \\
 &=&\frac{if_{P}}{4}{\int}_{0}^{1}dx \
    e^{i(xk{\cdot}z_{2}+\bar{x}k{\cdot}z_{1})} \Big\{
    k \!\!\!\slash {\gamma}_{5} {\Phi}_{P}(x)
    - {\mu}_{P} {\gamma}_{5} \Big[ {\Phi}_{P}^{p}(x)
    - {\sigma}_{{\mu}{\nu}}k^{\mu}z^{\nu}
       \frac{{\Phi}_{P}^{\sigma}(x)}{6} \Big] \Big\},
 \label{eq:LCDAs-01}
 \end{eqnarray}
 \begin{equation}
 \begin{array}{cl}
  \text{twist-2 asymptotic forms:} &
  {\Phi}_{P}(x)          =  6x\bar{x}, \\
  \text{twist-3 asymptotic forms:} &
  {\Phi}_{P}^{p}(x)      =  1, \ \ \ \ \
  {\Phi}_{P}^{\sigma}(x) =  6x\bar{x}.
 \end{array}
 \label{eq:LCDAs-02}
 \end{equation}
 where $f_{P}$ is a decay constant; $z=z_{2}-z_{1}$, $\bar{x}=1-x$.

 For a longitudinally polarized vector meson, we have
 \cite{0008255,9602323}:
 \begin{eqnarray}
 {\langle}0{\vert}\bar{q}(0){\gamma}_{\mu}q(z)
           {\vert}V(k,{\lambda}){\rangle} =
    k_{\mu} \frac{{\epsilon}^{\lambda}{\cdot}z}{k{\cdot}z}
    f_{V} m_{V} {\int}_{0}^{1} dx \
    e^{-ixk{\cdot}z} {\Phi}_{V}^{\|}(x),
 \label{eq:LCDAs-03}
 \end{eqnarray}
 \begin{equation}
 \text{twist-2  asymptotic forms:} \ \ {\Phi}_{V}^{\|}(x)=6x\bar{x}.
 \label{eq:LCDAs-04}
 \end{equation}
 where ${\epsilon}$ is a polarization vector, and
 ${\epsilon}_{\|}=k/m_{V}$.

 For the wave function of $B_{s}$ meson, we take
 \cite{0004004,0004213,bauer89}:
 \begin{equation}
 {\Phi}_{B}({\xi}) = N_{B} {\xi}^{2} (1-{\xi})^{2} {\exp} \Big[
  -\frac{ m_{B}^{2} {\xi}^{2} }{ 2 {\omega}_{B}^{2} }
  -\frac{ {\omega}_{B}^{2} b^{2} }{ 2 }  \Big],
 \label{eq:B-wave-function}
 \end{equation}
 where $N_{B}$ is the normalization constant. ${\Phi}_{B}({\xi})$ is
 peaked around ${\xi}$ ${\approx}$ 0.1 with ${\omega}_{B}=0.4\,\text{GeV}$
 and $b=0$.

 As to the divergent endpoint integral ${\int}dx/x$, in analogy with the
 treatment in works \cite{0104110}, we parameterize it as
 \begin{equation}
 X = {\int}^{1}_{0} \, \frac{dx}{x}
   = ( 1 + {\varrho} \, e^{i{\phi}} ) \, {\ln} \,
     \frac{m_{b}}{{\Lambda}_{h}},
   \ \ \ \ \ \ \ {\varrho} {\le} 1,
   \ \ \ \ \ \ \ 0^{\circ} {\le} {\phi} {\le} 360^{\circ}.
 \label{eq:x}
 \end{equation}
 In numerical calculation, we take their default values as
 \[ \begin{array}{c|c|c|c|c|c} \hline
  \text{decay modes} & {\varrho}_{H} & {\phi}_{H}
 & {\varrho}_{A} \,\text{\cite{0209233}} 
 & {\phi}_{A}    \,\text{\cite{0209233}}
 & {\Lambda}_{h} \,\text{\cite{0104110}} \\ \hline
 B_{s} {\to} PP & 0 & 0 & 0.5 &  10^{\circ} & 0.5\,\text{GeV} \\
 B_{s} {\to} PV & 0 & 0 & 1.0 & 330^{\circ} & 0.5\,\text{GeV} \\ \hline
 \end{array} \]
 where $({\varrho}_{H},{\phi}_{H})$ and $({\varrho}_{A},{\phi}_{A})$ are
 related to the contributions from hard spectator scattering and weak
 annihilations, respectively.

 \section{Branching ratios}
 \label{sec4}
 The branching ratios for charmless $B_{s}$ ${\to}$ $PP$, $PV$ decays in
 $B_{s}$ meson rest frame can be written as:
 \begin{equation}
  BR (B_{s}{\to}M_{1}M_{2})= \frac{{\tau}_{B_{s}}}{8{\pi}}
  \frac{{\vert}p{\vert}}{m_{B_{s}}^{2}}
        {\vert}{\cal A}(B_{s}{\to}M_{1}M_{2}){\vert}^{2}.
 \label{eq:br-01}
 \end{equation}
 where
 \begin{equation}
 {\vert}p{\vert}=\frac{\sqrt{[m_{B_{s}}^{2}-(m_{M_{1}}+m_{M_{2}})^{2}]
       [m_{B_{s}}^{2}-(m_{M_{1}}-m_{M_{2}})^{2}]}}{2m_{B_{s}}},
 \label{eq:br-02}
 \end{equation}
 The lifetime and mass for $B_{s}$ meson are
 ${\tau}_{B_{s}}$ = 1.461 ps, and
 $m_{B_{s}}$ = 5369.6 MeV \cite{pdg2002}.
 And since the QCDF approach works in the heavy quark limit, we take the
 masses of light mesons as zero in the computation of phase space, then
 ${\vert}p{\vert}=m_{B_{s}}/2$.

\begin{figure}[ht]
 \includegraphics[200,155][400,685]{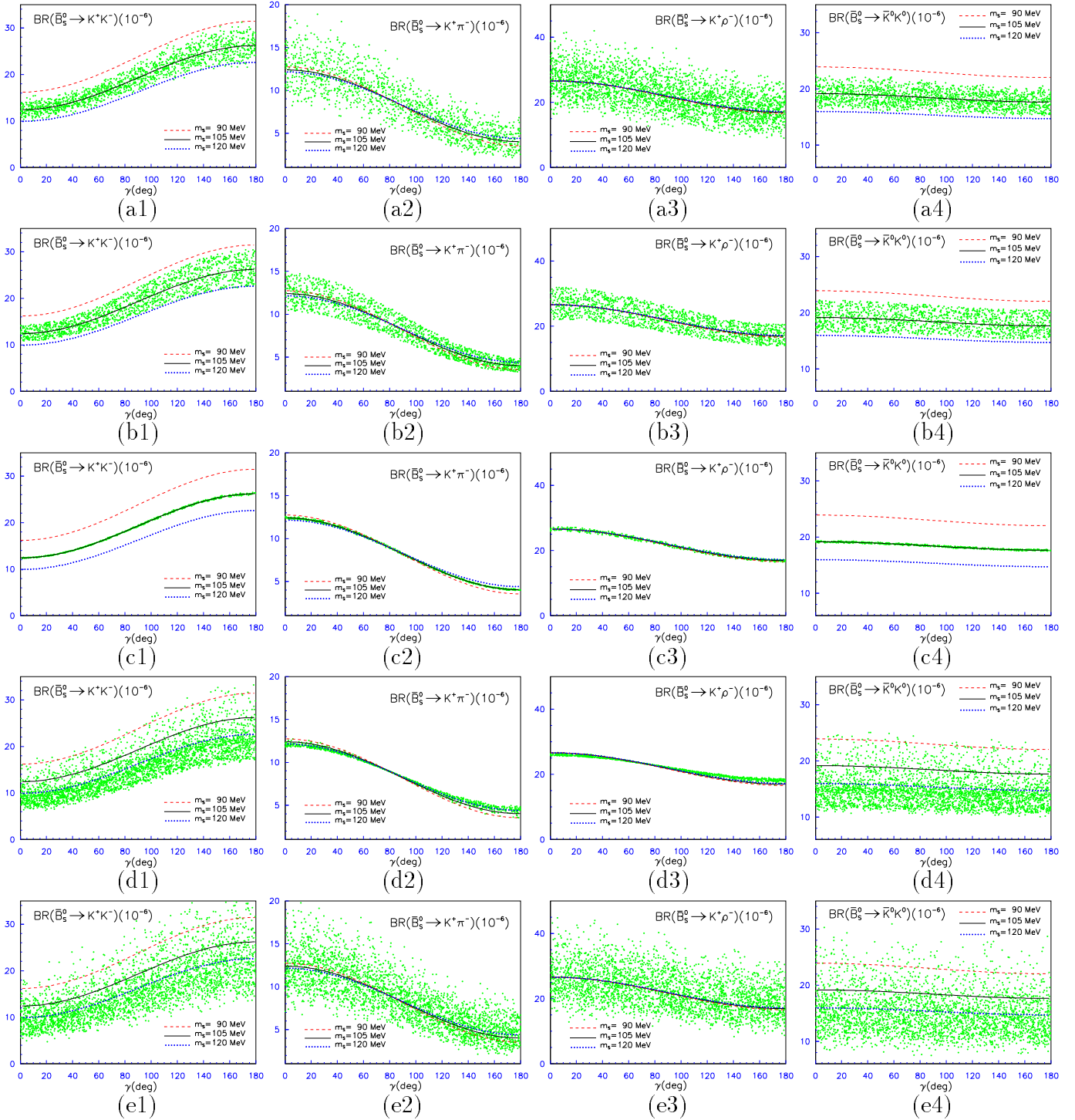}
 \caption{$CP$-averaged branching ratios of $B_{s}$ ${\to}$ $K^{+}K^{-}$
  (column 1), $K^{\pm}{\pi}^{\mp}$ (column 2), $K^{\pm}{\rho}^{\mp}$
  (column 3), and ${\overline{K}}^{0}K^{0}$ (column 4) decays within the
  QCDF approach versus the angle ${\gamma}$. The dashed lines, solid
  lines, and doted lines correspond to the default values of various
  theory inputs, wherein ${\overline{m}}_{s}(2\,\text{GeV})$ = 90 MeV,
  105 MeV, and 120 MeV, respectively. The dot-shades demonstrate the
  uncertainties due to the variations of the CKM matrix parameters $A$,
  ${\lambda}$, ${\bar{\rho}}$, ${\bar{\eta}}$ (row 1), formfactor
  $F_{0,1}^{B_{s}{\to}K}$ (row 2), (${\varrho}_{H}$, ${\phi}_{H}$)
  (row 3), (${\varrho}_{A}$, ${\phi}_{A}$) (row 4), and overall
  inputs (row 5).}
 \label{fig1}
 \end{figure}

\begin{figure}[ht]
 \includegraphics[235,670][335,770]{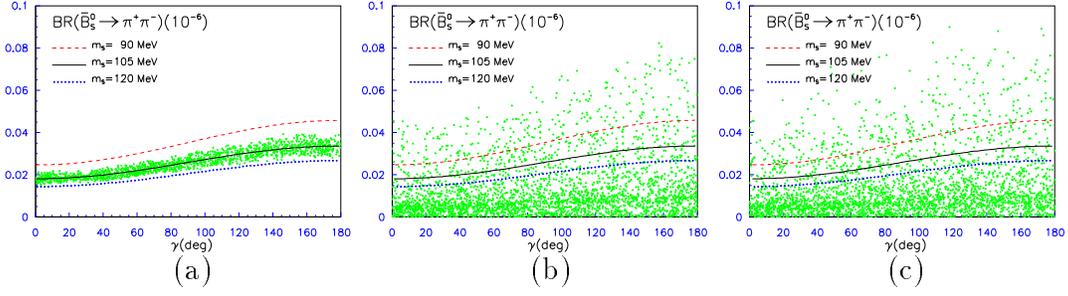}
 \caption{$CP$-averaged branching ratios of $B_{s}$ ${\to}$ ${\pi}^{+}{\pi}^{-}$
  decays within the QCDF approach versus the angle ${\gamma}$. The legends on
  the (dashed, solid, and doted) lines are the same as those in Fig. \ref{fig1}.
  The dot-shades demonstrate the uncertainties due to the variations of the CKM
  matrix parameters $A$, ${\lambda}$, ${\bar{\rho}}$, ${\bar{\eta}}$ (a), weak
  annihilation parameters (${\varrho}_{A}$, ${\phi}_{A}$) (b), and overall
  inputs (c).}
 \label{fig2}
 \end{figure}

 The numerical results of $CP$ averaged branching ratios for $B_{s}$
 decays are listed in Table \ref{tab4} and Table \ref{tab5}, which are
 calculated at the scale of ${\mu}$ = $m_{b}$ with two sets of CKM matrix
 parameters. The data in $BR$ columns are calculated within the NF
 framework and $BR$ ${\propto}$ ${\vert}{\cal A}^{f}{\vert}^{2}$; the
 data in $BR^{f}$ and $BR^{f+a}$ columns are estimated with the QCDF
 approach, and $BR^{f}$ ${\propto}$ ${\vert}{\cal A}^{f}{\vert}^{2}$,
 $BR^{f+a}$ ${\propto}$ ${\vert}{\cal A}^{f}+{\cal A}^{a}{\vert}^{2}$.
 In the following there are some remarks
 \begin{itemize}
 \item Only several interesting decay modes, such as
       $B_{s}$ ${\to}$ $K^{({\ast})}K$, $K^{({\ast}){\pm}}{\pi}^{\mp}$,
       $K^{\pm}{\rho}^{\mp}$, ${\eta}^{(\prime)}{\eta}^{(\prime)}$,
       have large branching ratios, which might be observed potentially
       in the near future. Branching ratios of other decay modes are
       small, not exceeding $1{\times}10^{-6}$. Especially for decays
       $B_{s}$ ${\to}$ ${\pi}{\eta}^{(\prime)}$, ${\pi}{\phi}$,
       ${\rho}{\eta}^{(\prime)}$, ${\omega}{\eta}^{(\prime)}$,
       whose tree contributions are suppressed by both CKM factor and
       color, and penguin contributions are electroweak coefficient
       $a_{9}$ dominant, so their $CP$ averaged branching ratios are very
       small, around ${\cal O}(10^{-7})$. As to these pure weak
       annihilation decays, such as $B_{s}$ ${\to}$ ${\pi}{\pi}$,
       ${\pi}{\rho}$, ${\pi}{\omega}$, their branching ratios are
       extremely small, around ${\cal O}(10^{-8})$.
 \item For those $b{\to}s$ transition decay modes, such as $B_{s}$
       ${\to}$ ${\eta}^{(\prime)}{\eta}^{(\prime)}$, $K^{({\ast})}K$,
       their tree contributions are CKM suppressed, and ``nonfactorizable''
       effects contribute a large portion to penguin coefficients $a_{4,6}$,
       so penguin contributions and tree ones are either competitive, or
       penguin dominant. So we can see large ``nonfactorizable'' effects
       in these decays. In addition, the coefficients $b_{1}$ and/or
       $b_{3}$ appear in these decay amplitudes, and the data in Tables
       \ref{tab4} and \ref{tab5} show that weak annihilation contributions
       are sizeable (${\gtrsim}50\%$). For those $b{\to}d$ transition decay
       modes, such as $B_{s}$ ${\to}$ $K^{(\ast){\pm}}{\pi}^{\mp}$,
       $K^{\pm}{\rho}^{\mp}$, they are $a_{1}$ dominant, and the radiative
       corrections are ${\alpha}_{s}$ suppressed compared with leading
       order contributions, so the data in Table \ref{tab4} and \ref{tab5}
       show no large difference between the results obtained with the QCDF
       approach and the NF's ones.
 \item There are hierarchy among some decay modes, such as
       \begin{eqnarray}
       & & BR({\overline{B}}_{s}^{0}{\to}K^{+}K^{-})
        >  BR({\overline{B}}_{s}^{0}{\to}K^{+}K^{{\ast}-})
        >  BR({\overline{B}}_{s}^{0}{\to}K^{-}K^{{\ast}+}),
        \label{eq:br-03} \\
       & & BR({\overline{B}}_{s}^{0}{\to}{\overline{K}}^{0}K^{0})
        >  BR({\overline{B}}_{s}^{0}{\to}K^{0}{\overline{K}}^{{\ast}0})
        >  BR({\overline{B}}_{s}^{0}{\to}{\overline{K}}^{0}K^{{\ast}0}),
        \label{eq:br-04} \\
       & & BR({\overline{B}}_{s}^{0}{\to}K^{+}{\pi}^{-})
        >  BR({\overline{B}}_{s}^{0}{\to}K^{{\ast}+}{\pi}^{-}),
        \label{eq:br-05}
       \end{eqnarray}
       There are two lines of reason for the above relations. One main line
       of reason is that the penguin contributions are important or/and
       dominant for these decays. Their decay amplitudes involve the QCD
       penguin parameters $a_{4}$ and $a_{6}$ in the form of $a_{4}$ +
       $Ra_{6}$, where $R>0$, for $B_{s}$ ${\to}$ $PP$ decays, and $R=0$
       ($R<0$) for $B_{s}$ ${\to}$ $PV$ decays with $B{\to}P$ ($B{\to}V$)
       transition, respectively, as stated in \cite{9809364}. The other
       line for the second inequality of Eq. (\ref{eq:br-03}) and
       Eq. (\ref{eq:br-04}) is $f_{K^{\ast}}F_{1}^{B_{s}^{0}{\to}K}$ $>$
       $f_{K}A_{0}^{B_{s}^{0}{\to}K^{\ast}}$. The numerical data in Tables
       \ref{tab4} and \ref{tab5} confirm the above relations in general.
       Here we would like to point out that because the weak annihilation
       parameters $b_{3}(P,V)=-b_{3}(V,P)$, and the combination of
       $b_{1}+b_{3}$ (or $b_{3}+2b_{4}$) is destructive for
       ${\overline{B}}_{s}^{0}$ ${\to}$ $K^{+}K^{{\ast}-}$
       (or $K^{0}{\overline{K}}^{{\ast}0}$), and constructive for
       ${\overline{B}}_{s}^{0}$ ${\to}$ $K^{-}K^{{\ast}+}$
       (or ${\overline{K}}^{0}K^{{\ast}0}$), the weak annihilation
       have more effects to decays of ${\overline{B}}_{s}^{0}$ ${\to}$
       $K^{-}K^{{\ast}+}$ (${\overline{K}}^{0}K^{{\ast}0}$) than to
       decays of ${\overline{B}}_{s}^{0}$ ${\to}$ $K^{+}K^{{\ast}-}$
       ($K^{0}{\overline{K}}^{{\ast}0}$), (for example, see the data in
       Table \ref{tab5}) which might be an explanation of why the data
       $BR^{f+a}({\overline{B}}_{s}^{0}{\to}K^{-}K^{{\ast}+})$ $>$
       $BR^{f+a}({\overline{B}}_{s}^{0}{\to}K^{+}K^{{\ast}-})$ in
       column 4 of Table \ref{tab5} that violate the second inequality
       of Eq. (\ref{eq:br-03}).
 \item It is interesting to note that
       $B_{s}$ ${\to}$ ${\eta}^{(\prime)}{\eta}^{(\prime)}$ decays
       have large branching ratios. In fact, their $SU(3)$ counterpart,
       $B_{u,d}$ ${\to}$ $K{\eta}^{\prime}$ have been reported to have
       the largest branching ratios among the two-body charmless rare
       $B$ decays,
       \[\begin{array}{c|c|c|c} \hline
          \text{Decay modes}
        & \text{CLEO  \cite{9912059}}
        & \text{BABAR \cite{0202022}}
        & \text{Belle \cite{0111037}} \\ \hline
       BR(B_{d}{\to}K_{s}{\eta}^{\prime}){\times}10^{6}   &
       89^{+18}_{-16}{\pm}9 & 42^{+13}_{-11}{\pm}4 &  55^{+19}_{-16}{\pm}8 \\
       BR(B_{u}{\to}K^{\pm}{\eta}^{\prime}){\times}10^{6} &
       80^{+10}_{-9}{\pm}7 & 70{\pm}8{\pm}5 & 79^{+12}_{-11}{\pm}9  \\ \hline
       \end{array}\]
       The abnormally large branching functions for $B_{u,d}$ ${\to}$
       $K{\eta}^{\prime}$ decays have triggered intense theoretical
       interests in understanding the special property of meson
       ${\eta}^{(\prime)}$. Several mechanisms have been proposed
       (for example, Refs \cite{9704357,9705304,9711428,9710509}).
       There are many works devoted to the study of exclusive $B$
       decays into two-body final states containing ${\eta}^{(\prime)}$,
       such as Refs. \cite{9807393,0210085}. It is generally believed
       that this problem is related to the axial anomaly in QCD, but
       the dynamical details remain unclear. It is now commonly believed
       that maybe there is large coupling between two gluons and
       ${\eta}^{\prime}$ which might have important contributions for
       ${\eta}^{\prime}$ production \cite{9711428,9710509}. It is shown
       in \cite{0012208} that the contributions of $g^{\ast}g^{\ast}$
       ${\to}$ ${\eta}^{\prime}$ to the  formfactors can give a good
       explanation for experimental data. Recently, M. Beneke and M.
       Neubert computed the exclusive $B_{u,d}$ ${\to}$
       ${\eta}^{(\prime)}+X$ decays using the QCDF approach \cite{0210085}.
       Their novel idea is to consider the flavor-singlet amplitudes for
       producing ${\eta}^{(\prime)}$ meson from not only a quark-antiquark
       pair but also a pair of gluons. Their analysis including three effects:
       $b{\to}sgg$ amplitude, spectator scattering involving two gluons,
       and weak annihilation, could qualitatively account for the
       measurements with inputs of specified values, but with large
       theoretical uncertainties. Their conclusion is that it is the
       constructive or destructive interference of non-singlet penguin
       amplitudes that is the key factor in explaining the exclusive
       $B$ ${\to}$ ${\eta}^{(\prime)}+X$ decays.
 \item From Table \ref{tab4} and \ref{tab5}, we can see that $CP$ averaged
       branching ratios for many decay modes are stable against the
       choices of the CKM matrix parameters. Only those decays which
       have large interference between tree contributions and penguin
       ones, such as $B_{s}$ ${\to}$ $K^{(\ast)}{\pi}$,
       $K^{(\ast){\pm}}K^{\mp}$, ${\cdots}$,
       are sensitive to the choice of the angle ${\gamma}$.
 \end{itemize}
 Of course, theoretical uncertainties from input parameters (such as
 the CKM matrix elements, quark masses, form factors, $X$, and so on)
 should be taken into account when discussing $B_{s}$ decays, which
 has been studied in detail in Refs. \cite{0104110,0108141,0201253}.
 In Figures \ref{fig1} and \ref{fig2}, we consider the effects of the
 variation of inputs on the $CP$ averaged branching ratios of $B_{s}$
 ${\to}$ $K^{+}K^{-}$, $K^{\pm}{\pi}^{\mp}$, $K^{\pm}{\rho}^{\mp}$,
 ${\overline{K}}^{0}K^{0}$ and ${\pi}^{+}{\pi}^{-}$ decays on the weak
 phase ${\gamma}$. In each plot, the dashed lines, solid lines, and
 doted lines give the QCDF's predictions with default values of various
 input parameters for ${\overline{m}}_{s}(2\,\text{GeV})$ = 90 MeV,
 105 MeV, and 120 MeV, respectively, but keep the ratio of light quark
 masses fixed, ${\overline{m}}_{q}/{\overline{m}}_{s}$ = 4.2/105.
 For discussion, we also vary the form factor $F_{0,1}^{B_{s}{\to}K}$
 by ${\pm}10\%$, i.e. $F_{0,1}^{B_{s}{\to}K}$ = 0.250 ${\sim}$ 0.300.
 It also includes the uncertainties from the CKM matrix parameters.
 From Figures \ref{fig1} and \ref{fig2}, we can see that
 \begin{itemize}
 \item There exist sizeable theoretical uncertainties which smear some
       helpful information on the angle ${\gamma}$ and demote the
       predictive power of the QCDF approach.
 \item The variation of form factors brings very large uncertainties (see
       the 2nd row of Figure \ref{fig1}) which in principle could be
       reduced by the ratios of branching ratios, while the uncertainties
       from $X_{H}$ which parameterizes the divergent end-point integral in
       hard spectator scattering corrections are very small (see the 3rd
       row of Figure \ref{fig1}).
 \item For tree dominant decay modes (such as $B_{s}$ ${\to}$
       $K^{\pm}{\pi}^{\mp}$, $K^{\pm}{\rho}^{\mp}$), the theoretical
       uncertainties mainly come from the formfactors and CKM matrix
       inputs. While for those decays with large interference between tree
       and penguin amplitudes, such as $B_{s}$ ${\to}$ $K^{+}K^{-}$,
       and penguin dominant decays, such as $B_{s}$ ${\to}$
       ${\overline{K}}^{0}K^{0}$, the theoretical uncertainties originate
       mainly from the variation of light quark masses and parameter
       $X_{A}$ besides formfactors.
 \item $B_{s}$ ${\to}$ ${\pi}^{+}{\pi}^{-}$ decay is pure annihilation
       process. Its amplitude is free from transition form factors and
       hard spectator scattering corrections. Hence the dominant
       theoretical uncertainties would come form the weak annihilation
       effects, more precisely the quantity $X_{A}$ (see Figure \ref{fig2}).
       Experimentally, it is worth searching for pure annihilation
       processes which may be helpful to learn more about the annihilation
       mechanism and to provide some useful information about final states
       interactions and nonperturbative parameters, such as $X_{A}$.
 \end{itemize}

 \section{$CP$ asymmetries}
 \label{sec5}
 $CP$-violating asymmetries for $B_{s}$ decays has been studies in
 \cite{9503278}. In this paper, we shall evaluate them with the QCDF
 approach. In principle, the calculation of $CP$-violating asymmetries
 for $B_{s}$ are similar with those for $B_{d}$ decays.
 Due to flavor-changing interactions, ${\overline{B}}_{s}^{0}$ and
 $B_{s}^{0}$ can oscillate into each other with time evolution. The
 time dependent $CP$ asymmetries ${\cal A}_{CP}$ for $B_{s}$ decays
 is defined as
 \begin{equation}
 {\cal A}_{CP}(t)=\frac{{\Gamma}({\overline{B}}_{s}^{0}(t){\to}\bar{f})
                       -{\Gamma}(B_{s}^{0}(t){\to}f)}
                       {{\Gamma}({\overline{B}}_{s}^{0}(t){\to}\bar{f})
                       +{\Gamma}(B_{s}^{0}(t){\to}f)}.
 \label{eq:cp-01}
 \end{equation}
 As discussed previously in Refs. \cite{0108141,0201253}, the $B_{s}$
 ${\to}$ $PP$, $PV$ decays can be classified into three cases according
 to the properties of the final states,
 \begin{itemize}
 \item case-I: $B_{s}^{0}$ ${\to}$ $f$, ${\overline{B}}_{s}^{0}$ ${\to}$
       $\bar{f}$, but $B^{0}_{s}$ ${\not\to}$ $\bar{f}$,
       ${\overline{B}}_{s}^{0}$ ${\not\to}$ $f$, for example,
       ${\overline{B}}_{s}^{0}$ ${\to}$ $K^{+}{\rho}^{-}$,
       ${\pi}^{-}K^{{\ast}+}$, ${\cdots}$, the $CP$-violating asymmetry
       for these decays is time independent,
       \begin{equation}
       {\cal A}_{CP}=\frac{{\Gamma}({\overline{B}}_{s}^{0}{\to}\bar{f})
                          -{\Gamma}(B_{s}^{0}{\to}f)}
                          {{\Gamma}({\overline{B}}_{s}^{0}{\to}\bar{f})
                          +{\Gamma}(B_{s}^{0}{\to}f)}.
 \label{eq:cp-02}
 \end{equation}
 \item case-II: $B_{s}^{0}$ ${\rightarrow}$ $(f=\bar{f})$ ${\leftarrow}$
       ${\overline{B}}_{s}^{0}$, for example, $B_{s}$ ${\to}$
       $K^{\pm}K^{\mp}$, ${\eta}^{(\prime)}{\eta}^{(\prime)}$,
       ${\cdots}$, the time-integrated $CP$-violating asymmetry for
       these decays is
       \begin{equation}
       {\cal A}_{CP}=
          \frac{1}{1+x_{s}^{2}}a_{{\epsilon}^{\prime}}
         +\frac{x_{s}}{1+x_{s}^{2}}a_{{\epsilon}+{\epsilon}^{\prime}},
       \label{eq:cp-03}
       \end{equation}
       \begin{equation}
       a_{{\epsilon}^{\prime}}=
           \frac{1-{\vert}{\lambda}_{CP}{\vert}^{2}}
                {1+{\vert}{\lambda}_{CP}{\vert}^{2}},\ \ \
       a_{{\epsilon}+{\epsilon}^{\prime}}=
           \frac{-2\,\text{Im}({\lambda}_{CP})}
                {1+{\vert}{\lambda}_{CP}{\vert}^{2}},\ \ \
       {\lambda}_{CP}=
          \frac{V_{ts}V_{tb}^{\ast}}{V_{ts}^{\ast}V_{tb}}\;
          \frac{{\cal A}({\overline{B}}_{s}^{0}(0){\to}\bar{f})}
               {{\cal A}(B_{s}^{0}(0){\to}f)},
       \label{eq:cp-04}
       \end{equation}
      where $a_{{\epsilon}^{\prime}}$ and
      $a_{{\epsilon}+{\epsilon}^{\prime}}$ are direct and mixing-induced
      $CP$-violating asymmetries, respectively. The parameter $x_{s}$ =
      ${\Delta}m_{B_{s}}/{\Gamma}_{B_{s}}$ is considerably large for
      $B_{s}$ system, $x_{s}$ $>19.0$ at $95\%$ C.L. \cite{pdg2002}. In
      our calculation, we shall take the preferred value in the SM $x_{s}$
      ${\simeq}$ 20 \cite{9408332}. Clearly, $CP$-violating asymmetry
      ${\cal A}_{CP}$ should be very small because
      $a_{{\epsilon}^{\prime}}$ and $a_{{\epsilon}+{\epsilon}^{\prime}}$
      in Eq. (\ref{eq:cp-03}) are strongly suppressed by $1/x_{s}^{2}$ and
      $1/x_{s}$, respectively.
 \item case-III: $B_{s}^{0}$ ${\rightarrow}$ $(f\&\bar{f})$ ${\leftarrow}$
      ${\overline{B}}_{s}^{0}$, for example, $B_{s}$ ${\to}$
      $(K_{S}^{0}{\overline{K}}^{{\ast}0}$ ${\&}$ $K_{S}^{0}K^{{\ast}0})$,
      $(K^{+}K^{{\ast}-}$ ${\&}$ $K^{-}K^{{\ast}+})$,
      $({\pi}^{+}{\rho}^{-}$ ${\&}$ ${\pi}^{-}{\rho}^{+})$.
      Analogous to the notations for $B_{d}$ decays in \cite{9804363},
      the time dependent decay widths for this case of $B_{s}$ decays are
      written as:
      \begin{eqnarray}
      & &{\Gamma}(B_{s}^{0}(t){\to}f)=
          \frac{e^{-{\Gamma}_{B_{s}}t}}{2}
          \Big( {\vert}g{\vert}^{2}+{\vert}h{\vert}^{2} \Big)
          \Big[1+a_{{\epsilon}^{\prime}}{\cos}({\Delta}m_{B_{s}}t)
                +a_{{\epsilon}+{\epsilon}^{\prime}}
                    {\sin}({\Delta}m_{B_{s}}t) \Big],
      \label{eq:cp-05} \\
      & &{\Gamma}({\overline{B}}^{0}(t){\to}\bar{f})=
          \frac{e^{-{\Gamma}_{B_{s}}t}}{2}
          \Big( {\vert}\bar{g}{\vert}^{2}+{\vert}\bar{h}{\vert}^{2} \Big)
          \Big[1-a_{{\overline{\epsilon}}^{\prime}}
                    {\cos}({\Delta}m_{B_{s}}t)
                -a_{{\epsilon}+{\overline{\epsilon}}^{\prime}}
                    {\sin}({\Delta}m_{B_{s}}t) \Big],
      \label{eq:cp-06} \\
      & &{\Gamma}(B^{0}(t){\to}\bar{f})=
          \frac{e^{-{\Gamma}_{B_{s}}t}}{2}
          \Big( {\vert}\bar{g}{\vert}^{2}+{\vert}\bar{h}{\vert}^{2} \Big)
          \Big[1+a_{{\overline{\epsilon}}^{\prime}}
                    {\cos}({\Delta}m_{B_{s}}t)
                +a_{{\epsilon}+{\overline{\epsilon}}^{\prime}}
                    {\sin}({\Delta}m_{B_{s}}t) \Big],
      \label{eq:cp-07} \\
      & &{\Gamma}({\overline{B}}^{0}(t){\to}f)=
          \frac{e^{-{\Gamma}_{B_{s}}t}}{2}
          \Big( {\vert}g{\vert}^{2}+{\vert}h{\vert}^{2} \Big)
          \Big[1-a_{{\epsilon}^{\prime}}{\cos}({\Delta}m_{B_{s}}t)
                -a_{{\epsilon}+{\epsilon}^{\prime}}
                    {\sin}({\Delta}m_{B_{s}}t) \Big],
      \label{eq:cp-08}
      \end{eqnarray}
      where
      \begin{eqnarray}
       & & g ={\cal A}(B_{s}^{0}(0){\to}f),
      \ \ \ \ \ \ \ \ \ \
      \bar{g}={\cal A}({\overline{B}}_{s}^{0}(0){\to}\bar{f}),
      \label{eq:cp-09} \\
       & & h ={\cal A}({\overline{B}}_{s}^{0}(0){\to}f),
      \ \ \ \ \ \ \ \ \ \
      \bar{h}={\cal A}(B_{s}^{0}(0){\to}\bar{f}),
      \label{eq:cp-10}
      \end{eqnarray}
      and with $q/p$ = $V_{ts}V_{tb}^{\ast}/V_{ts}^{\ast}V_{tb}$,
      \begin{eqnarray}
   & & a_{{\epsilon}^{\prime}}
          =\frac{{\vert}g{\vert}^{2}-{\vert}h{\vert}^{2}}
                {{\vert}g{\vert}^{2}+{\vert}h{\vert}^{2}},
       \ \ \ \ \ \ \ \ \
       a_{{\epsilon}+{\epsilon}^{\prime}}
          =\frac{-2\,\text{Im}[(q/p){\times}(h/g)]}
                {1+{\vert}h/g{\vert}^{2}},
       \label{eq:cp-11} \\
   & & a_{{\overline{\epsilon}}^{\prime}}
          =\frac{{\vert}\bar{h}{\vert}^{2}-{\vert}\bar{g}{\vert}^{2}}
                {{\vert}\bar{h}{\vert}^{2}+{\vert}\bar{g}{\vert}^{2}},
       \ \ \ \ \ \ \ \ \
       a_{{\epsilon}+{\overline{\epsilon}}^{\prime}}
          =\frac{-2\,\text{Im}[(q/p){\times}({\bar{g}}/{\bar{h}})]}
                {1+{\vert}{\bar{g}}/{\bar{h}}{\vert}^{2}}.
       \label{eq:cp-12}
       \end{eqnarray}
 \end{itemize}

 Our numerical results of $CP$-violating asymmetries for $B_{s}$ ${\to}$
 $PP$, $PV$ decays are listed in Tables \ref{tab6}--\ref{tab9}, which
 are calculated with two sets of CKM matrix parameters. In the following
 there are some remarks.
 \begin{itemize}
 \item From Tables \ref{tab8}--\ref{tab9}, we can see that as expected,
       due to the large parameter $x_{s}$ suppression, $CP$-violating
       asymmetry ${\cal A}_{CP}$ for these case-II decay modes are indeed
       very small, not exceeding $5\%$.
 \item From the QCDF formula Eq. (\ref{eq:qcdf}), we know that radiative
       corrections and ``nonfactorizable'' contributions should be
       either at the order of ${\alpha}_{s}$ or power suppressed in
       ${\Lambda}_{QCD}/m_{b}$, therefore, the $CP$-violating asymmetries
       ${\cal A}_{CP}$ for these $a_{1}$ dominant decays, such as $B_{s}$
       ${\to}$ $K^{(\ast){\pm}}{\pi}^{\mp}$, $K^{\pm}{\rho}^{\mp}$, are
       not large because of the small strong phases.
 \item From the previous $B_{u,d}$ ${\to}$ $PP$, $PV$ analysis
       \cite{0108141,0201253}, we know that ``nonfactorizable'' effects
       contribute a large imaginary part to the coefficients $a_{2,4,6}$.
       So from Tables \ref{tab6}--\ref{tab9}, we see that for these decays
       which have large interference between tree contributions dominated
       by $a_{2}$ and QCD penguin ones, and large interference between
       CKM suppressed $a_{2}$ dominant tree contributions and electroweak
       penguin ones, there exist large direct $CP$-violating asymmetries,
       for example,  ${\cal A}_{CP}$ for ${\overline{B}}_{s}^{0}$ ${\to}$
       $K^{{\ast}0}{\pi}^{0}$, $K^{{\ast}0}{\eta}^{(\prime)}$ decays,
       $a_{{\epsilon}^{\prime}}$ for $B_{s}$ ${\to}$ $K_{S}^{0}{\pi}^{0}$,
       $K_{S}^{0}{\eta}^{(\prime)}$, $K_{S}^{0}{\rho}^{0}$,
       $K_{S}^{0}{\omega}$ decays, and
       $a_{{\epsilon}^{\prime}}$ for $B_{s}$ ${\to}$ ${\pi}^{0}{\phi}$,
       ${\pi}^{0}{\eta}^{(\prime)}$, ${\eta}^{(\prime)}{\rho}^{0}$,
       ${\eta}^{(\prime)}{\omega}$ decays.
 \item The $CP$-violating asymmetries for the most decay modes is not
       keen on the variation of the CKM matrix parameters. The large
       and direct $CP$-violating asymmetries which are sensitive to the
       angle ${\gamma}$ are only display for some of these decays, such
       as ${\cal A}_{CP}$ for ${\overline{B}}_{s}^{0}$ ${\to}$
       $K^{{\ast}0}{\pi}^{0}$, $K^{{\ast}0}{\eta}^{(\prime)}$, and
       $a_{{\epsilon}^{\prime}}$ for $B_{s}$ ${\to}$ $K_{S}^{0}{\pi}^{0}$,
       $K_{S}^{0}{\omega}$,
 \item If we assume that $B_{s}^{0}$-${\overline{B}}_{s}^{0}$ mixing phase
       is negligible, or $q/p$ = 1, it should be convenient to determine
       the weak angle ${\gamma}$ or ${\beta}$ from measurements of
       $CP$-violating
       asymmetries for $B_{s}$ decays. Unfortunately, the very rapid
       $B_{s}^{0}$-${\overline{B}}_{s}^{0}$ oscillations are expected
       due to the large mixing parameter $x_{s}$ in SM, which makes the
       experimental studies of $CP$ violation in $B_{s}$ meson system
       difficult, and only bounds on few decays modes are given for the
       moment. In addition, there exist large theoretical uncertainties.
       Within the QCDF approach, the subleading power corrections in
       $1/m_{b}$ are might be as important as the radiative corrections
       numerically because $m_{b}$ is not infinitely large. So the QCDF
       approach can only give the order of magnitude of the $CP$-violating
       asymmetries,  as stated in \cite{0201253}.
 \end{itemize}

 \section{Extracting weak phases from $B_{s}{\to}KK$ decays}
 \label{sec6}
 It is necessary to test the self-consistency of the CKM description of
 $CP$ violation through a variety of processes. One test involves the
 $B_{d}(t)$ ${\to}$ ${\pi}^{+}{\pi}^{-}$ decays which are potentially
 rich sources of information of both strong and weak phases. Experimentally,
 BABAR and Belle have reported the measurements of $CP$-violating
 asymmetries in $B_{d}(t)$ ${\to}$ ${\pi}^{+}{\pi}^{-}$ decays,
 \[\begin{array}{c|c|c|c} \hline
    & \text{BABAR \cite{0207055}} & \text{Belle \cite{0204002}}
    & \text{Average} \\ \hline
  S_{{\pi}{\pi}} & -0.02{\pm}0.34{\pm}0.05
                 &  1.21_{-0.38-0.16}^{+0.27+0.13}
                 &  0.48{\pm}0.35 \footnotemark[5] \\ \hline
  C_{{\pi}{\pi}} & -0.30{\pm}0.25{\pm}0.04
                 & -0.94_{-0.25}^{+0.31}{\pm}0.09
                 & -0.54{\pm}0.20 \\ \hline
   \end{array}\]
 \footnotetext[5]{the error includes a scale factor of 1.3}
 which has triggered high theoretical interest. Theoretically, if we
 assume that the penguin amplitudes are zero for $B_{d}(t)$ ${\to}$
 ${\pi}^{+}{\pi}^{-}$ decays, then it is expected to determine the weak
 angle ${\alpha}$ from the $S_{{\pi}{\pi}}$ = $-{\sin}2{\alpha}$.
 Unfortunately, this relation is strongly polluted by penguin effects.
 The nature of tree and penguin amplitudes lead to some indeterminacy in
 determining the weak and strong phases. Hence, it will be an interesting
 work to investigate the ratio of penguin to tree amplitudes. Using the
 $U$-spin symmetry, we can get some additional information on the
 penguin-to-tree ratio, $P_{{\pi}{\pi}}/T_{{\pi}{\pi}}$, from its
 counterpart $B_{s}$ ${\to}$ $K^{+}K^{-}$ as a cross check. To illustrate,
 we now describe the expressions for the decay amplitudes of $B_{s}$
 ${\to}$ $K^{+}K^{-}$ and $B_{d}$ ${\to}$ ${\pi}^{+}{\pi}^{-}$ as follows:
 \begin{eqnarray}
 {\cal A}({\overline{B}}^{0}{\to}{\pi}^{+}{\pi}^{-}) &=&
 {\vert}T_{{\pi}{\pi}}{\vert} e^{-i{\delta}_{T}} e^{-i{\gamma}} +
 {\vert}P_{{\pi}{\pi}}{\vert} e^{-i{\delta}_{P}},
 \label{eq:pipi} \\
 {\cal A}({\overline{B}}^{0}_{s}{\to}K^{+}K^{-}) &=&
 {\vert}T_{c}{\vert} e^{-i{\delta}_{T}^{c}} e^{-i{\gamma}} \ \, -
 {\vert}P_{c}{\vert} e^{-i{\delta}_{P}^{c}},
 \label{eq:kk-1}
 \end{eqnarray}
 Using the $SU(3)$ flavor symmetry, we have \cite{0203158}
 \begin{eqnarray}
 \frac{{\vert}T_{{\pi}{\pi}}{\vert}}{{\vert}T_{c}{\vert}} &=&
 \frac{{\vert}V_{ub}V_{ud}^{\ast}{\vert}}
      {{\vert}V_{ub}V_{us}^{\ast}{\vert}} =
 \frac{1-{\lambda}^{2}/2}{\lambda}, \ \ \ \ \ \ \ \
 {\delta}_{T} = {\delta}_{T}^{c},
 \label{eq:tt} \\
 \frac{{\vert}P_{{\pi}{\pi}}{\vert}}{{\vert}P_{c}{\vert}} &=&
 \frac{{\vert}V_{cb}V_{cd}^{\ast}{\vert}}
      {{\vert}V_{cb}V_{cs}^{\ast}{\vert}} =
 \frac{\lambda}{1-{\lambda}^{2}/2}, \ \ \ \ \ \ \ \
 {\delta}_{P} = {\delta}_{P}^{c},
 \label{eq:pp} \\
 \frac{{\vert}P_{{\pi}{\pi}}{\vert}}{{\vert}T_{{\pi}{\pi}}{\vert}}
  &=&
 {\tan}^{2} {\theta}_{c} \frac{{\vert}P_{c}{\vert}}{{\vert}T_{c}{\vert}},
 \ \ \ \ \ \ \
 {\delta}_{T}-{\delta}_{P}={\delta}_{T}^{c}-{\delta}_{P}^{c}
  ={\delta}^{\prime},
 \label{eq:tp}
 \end{eqnarray}
 and with $r_{A}$ = $\frac{f_{B_{s}}f_{K}}{F^{B_{s}{\to}K}m_{B_{s}}^{2}}$,
 \begin{equation}
 \frac{P_{c}}{T_{c}} =
 \frac{- {\vert}V_{cb}V_{cs}^{\ast}{\vert} \{ a_{4}^{c}+a_{10}^{c}
  +r_{\chi}(a_{6}^{c}+a_{8}^{c})+r_{A}(b_{3}+2b_{4}
 -\frac{1}{2}b_{3}^{ew}+\frac{1}{2}b_{4}^{ew}) \} }
 {{\vert}V_{ub}V_{us}^{\ast}{\vert} \{ a_{1}^{u}+a_{4}^{u}+a_{10}^{u}
  +r_{\chi}(a_{6}^{u}+a_{8}^{u})+r_{A}(b_{1}+b_{3}+2b_{4}
 -\frac{1}{2}b_{3}^{ew}+\frac{1}{2}b_{4}^{ew}) \} },
 \label{eq:kk-tp}
 \end{equation}
 Compared with $B_{d}$ ${\to}$ ${\pi}^{+}{\pi}^{-}$ decay, the
 contribution of $T_{c}$ ($P_{c}$) for $B_{s}$ ${\to}$ $K^{+}K^{-}$
 decay is reduced (enhanced) by ${\tan}\,{\theta}_{c}$. Of course,
 the relations of Eq. (\ref{eq:tt}) and Eq. (\ref{eq:pp}) are affected
 by $U$-spin breaking effects, such as factor
 $\frac{(m_{B_{d}}^{2}-m_{\pi}^{2})F^{B{\to}{\pi}}f_{\pi}}
       {(m_{B_{s}}^{2}-m_{K}^{2})F^{B_{s}{\to}K}f_{K}}$, and so on.
 But $SU(3)$ flavor breaking effects are expected to be very small
 in Eq. (\ref{eq:tp}) within factorization approach.
 As stated in \cite{0011323,0003238}, assuming that
 $B_{s}^{0}$-${\overline{B}}_{s}^{0}$ mixing phase is negligible,
 and taking the angle ${\beta}$ as one known input which can be
 determined from $B_{d}$ ${\to}$ $J/{\Psi}K_{S}$ decays and has
 been tentatively given by BABAR \cite{0203007} and Belle \cite{0202027},
 the strong phase ${\delta}_{T}-{\delta}_{P}$ and
 ${\vert}P_{{\pi}{\pi}}/T_{{\pi}{\pi}}{\vert}$
 (or ${\delta}_{T}^{c}-{\delta}_{P}^{c}$ and ${\vert}P_{c}/T_{c}{\vert}$)
 as a function of weak  angle ${\gamma}$ or/and ${\alpha}$ can be
 determined from measurements of $S_{{\pi}{\pi}}$ and $C_{{\pi}{\pi}}$
 (or $S_{KK}$ and $C_{KK}$). And employing the relation Eq. (\ref{eq:tp})
 within the $U$-spin symmetry, the penguin-to-tree ratio  can be
 overconstrained from $B_{d}$ ${\to}$ ${\pi}^{+}{\pi}^{-}$ and $B_{s}$
 ${\to}$ $K^{+}K^{-}$ decays. And using the value of the penguin-to-tree
 ratios, some information on weak phases ${\gamma}$ or/and ${\alpha}$
 can be extracted from the measurements,
 \begin{eqnarray}
  {\lambda}_{{\pi}{\pi}}&=&
    \frac{V_{td}V_{tb}^{\ast}}{V_{td}^{\ast}V_{tb}}
    \frac{{\cal A}({\overline{B}}^{0}(0){\to}{\pi}^{+}{\pi}^{-})}
         {{\cal A}(B^{0}(0){\to}{\pi}^{+}{\pi}^{-})} = e^{i2{\alpha}}
    \frac{1+{\vert}P_{{\pi}{\pi}}/T_{{\pi}{\pi}}{\vert}
             e^{i{\delta}^{\prime}}e^{i{\gamma}}}
         {1+{\vert}P_{{\pi}{\pi}}/T_{{\pi}{\pi}}{\vert}
             e^{i{\delta}^{\prime}}e^{-i{\gamma}}},
   \label{eq:lcppipi} \\
   S_{{\pi}{\pi}}&=&
     \frac{-2\,{\text{Im}}({\lambda}_{{\pi}{\pi}})}
          {1+{\vert}{\lambda}_{{\pi}{\pi}}{\vert}^{2}} =
     \frac{ -{\sin}2{\alpha}
            +2{\vert}P_{{\pi}{\pi}}/T_{{\pi}{\pi}}{\vert}
              {\cos}{\delta}^{\prime}{\cos}({\alpha}-{\beta})
            + {\vert}P_{{\pi}{\pi}}/T_{{\pi}{\pi}}{\vert}^{2}
              {\sin}2{\beta}}
          {1-2{\vert}P_{{\pi}{\pi}}/T_{{\pi}{\pi}}{\vert}
              {\cos}{\delta}^{\prime}{\cos}({\alpha}+{\beta})
            +{\vert}P_{{\pi}{\pi}}/T_{{\pi}{\pi}}{\vert}^{2}},
   \label{eq:scppipi} \\
   C_{{\pi}{\pi}}&=&
     \frac{1-{\vert}{\lambda}_{{\pi}{\pi}}{\vert}^{2}}
          {1+{\vert}{\lambda}_{{\pi}{\pi}}{\vert}^{2}} =
     \frac{  2{\vert}P_{{\pi}{\pi}}/T_{{\pi}{\pi}}{\vert}
             {\sin}{\delta}^{\prime}{\sin}{\gamma}}
          {1+2{\vert}P_{{\pi}{\pi}}/T_{{\pi}{\pi}}{\vert}
             {\cos}{\delta}^{\prime}{\cos}{\gamma}
            +{\vert}P_{{\pi}{\pi}}/T_{{\pi}{\pi}}{\vert}^{2}},
   \label{eq:ccppipi} \\
  {\lambda}_{KK}&=&
    \frac{V_{ts}V_{tb}^{\ast}}{V_{ts}^{\ast}V_{tb}}
    \frac{{\cal A}({\overline{B}}_{s}^{0}(0){\to}K^{+}K^{-})}
         {{\cal A}(B_{s}^{0}(0){\to}K^{+}K^{-})} = e^{-i2{\gamma}}
    \frac{1-{\vert}P_{c}/T_{c}{\vert}
          e^{i{\delta}^{\prime}}e^{i{\gamma}}}
         {1-{\vert}P_{c}/T_{c}{\vert}
          e^{i{\delta}^{\prime}}e^{-i{\gamma}}},
   \label{eq:lcpkk} \\
   S_{KK} &=& \frac{-2\,{\text{Im}}({\lambda}_{KK})}
                   {1+{\vert}{\lambda}_{KK}{\vert}^{2}} =
     \frac{{\sin}2{\gamma}-2{\vert}P_{c}/T_{c}{\vert}
              {\cos}{\delta}^{\prime}{\sin}{\gamma}}
          {1-2{\vert}P_{c}/T_{c}{\vert}
              {\cos}{\delta}^{\prime}{\cos}{\gamma}
            + {\vert}P_{c}/T_{c}{\vert}^{2}},
   \label{eq:scpkk} \\
   C_{KK} &=& \frac{1-{\vert}{\lambda}_{KK}{\vert}^{2}}
                   {1+{\vert}{\lambda}_{KK}{\vert}^{2}} =
     \frac{ -2{\vert}P_{c}/T_{c}{\vert}
              {\sin}{\delta}^{\prime}{\sin}{\gamma}}
          {1-2{\vert}P_{c}/T_{c}{\vert}
              {\cos}{\delta}^{\prime}{\cos}{\gamma}
            + {\vert}P_{c}/T_{c}{\vert}^{2}}.
   \label{eq:ccpkk}
 \end{eqnarray}

\begin{figure}[ht]
 \includegraphics[260,670][360,650]{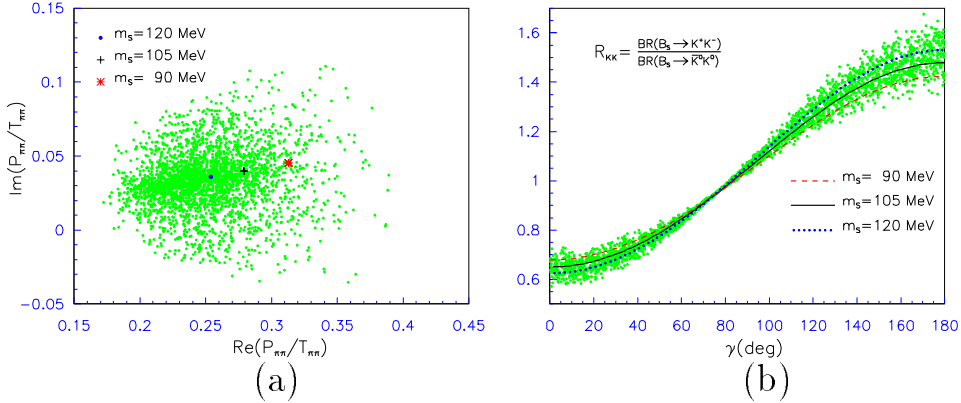}
 \caption{Penguin-to-tree ratio ${\vert}P_{{\pi}{\pi}}/T_{{\pi}{\pi}}{\vert}$
  (a) and $R_{KK}$ versus ${\gamma}$. The legends on the (dashed, solid, and
  doted) lines are the same as those in Fig. \ref{fig1}. The dot-shades
  demonstrate the uncertainties due to the variations of overall inputs.}
 \label{fig3}
 \end{figure}

 Here the numerical values of penguin-to-tree ratios are given in Table
 \ref{tab10}. Although using different derivation and inputs, within
 the range of one ${\sigma}$, the results of penguin-to-tree ratio
 ${\vert}P_{{\pi}{\pi}}/T_{{\pi}{\pi}}{\vert}$ that are calculated
 with Eq. (\ref{eq:tp}) are in agreement with the result of (28.5
 ${\pm}$ 5.1 ${\pm}$ 5.7) \% \cite{0104110} which is calculated
 with $X_{A}$ = $X_{H}$ = ${\ln}(m_{b}/{\Lambda}_{h})$, and the
 result of (27.6 ${\pm}$ 6.4) \% \cite{0109238} (including $SU(3)$
 breaking effects), but not as large as 0.41 \cite{0209233}. In
 addition, the value of the strong phase ${\delta}^{\prime}$ is also
 consistent with $(8.2{\pm}3.8)^{\circ}$ \cite{0104110}. The
 uncertainties of penguin-to-tree ratio $P_{{\pi}{\pi}}/T_{{\pi}{\pi}}$
 are shown in Figure \ref{fig3}, which indicates ${\delta}^{\prime}$
 ${\in}$ $(-7^{\circ},22^{\circ})$.

 In addition, if the penguin-to-tree ratios are determined,
 then it is expected to extract some information or give bound on weak
 angle ${\gamma}$ from the measurements of $B_{s}$ ${\to}$ $K^{+}K^{-}$,
 ${\overline{K}}^{0}K^{0}$ decays in the future. Now let's illustrate
 this point. The QCD penguin terms for these two decays should be equal
 according to factorization, so the decay amplitude for $B_{s}$ ${\to}$
 ${\overline{K}}^{0}K^{0}$ could be written as,
 \begin{equation}
 {\cal A}({\overline{B}}^{0}_{s}{\to}{\overline{K}}^{0}K^{0}) \,
 {\simeq} \, -{\vert}P_{c}{\vert} e^{-i{\delta}_{P}^{c}},
 \label{eq:kk-2}
 \end{equation}
 There are two approximations in Eq. (\ref{eq:kk-2}), as stated in
 \cite{0203158}: (1) The color suppressed terms which are proportional
 to the CKM matrix factor $V_{ub}V_{us}^{\ast}$ in Eq. (\ref{eq:app1-01})
 and Eq. (\ref{eq:app3-01}) are safely neglected because of
 ${\vert}V_{ub}V_{us}^{\ast}/V_{cb}V_{cs}^{\ast}{\vert}$ ${\simeq}$ 2\%.
 (2) The tiny isospin breaking effects are disregarded, and the small
 difference of electroweak penguin contributions between these two decay
 modes are neglected. The ratio of $CP$ averaged branching ratios is
 defined as
 \begin{eqnarray}
 R_{KK} &=& \frac{BR({\overline{B}}^{0}_{s}{\to}K^{+}K^{-})
                 +BR(B^{0}_{s}{\to}K^{+}K^{-})}
  {BR({\overline{B}}^{0}_{s}{\to}{\overline{K}}^{0}K^{0})
  +BR(B^{0}_{s}{\to}{\overline{K}}^{0}K^{0})} \nonumber \\
 &=& \frac{1-2{\vert}P_{c}/T_{c}{\vert}
             {\cos}{\delta}^{\prime}{\cos}{\gamma}
            +{\vert}P_{c}/T_{c}{\vert}^{2}}
          {{\vert}P_{c}/T_{c}{\vert}^{2}},
 \label{eq:rkk}
 \end{eqnarray}
 Then we can get a constraint on ${\gamma}$,
 \begin{equation}
 {\cos}{\gamma} {\gtrsim}
 \left\vert \frac{P_{c}}{T_{c}} \right\vert
 \left(1-\sqrt{R_{KK}}\right),
 \label{eq:cosr}
 \end{equation}
 In addition, Gronau and Ronsner also gave a bound on ${\gamma}$ from
 the decays $B_{s}$ ${\to}$ $K^{+}K^{-}$, ${\overline{K}}^{0}K^{0}$ without
 prior knowledge of the penguin-to-tree ratio, ${\sin}^{2}{\gamma}$ ${\le}$
 $R_{KK}$ \cite{0203158}.

 From the above discussion, we can see that sufficient measurements of
 $B_{s}$ decays in the future can resolve ambiguities on the determination
 of the CKM angles. Here, we estimate the bounds on ${\gamma}$ with the QCDF
 approach. The data in Table \ref{tab10} indicates that the weak angle
 ${\gamma}$ given in Refs. \cite{0207101,0209233} and the bound from
 Eq. (\ref{eq:cosr}) are consistent with each other, which might be tested
 in future measurements.

 \section{Summary and Conclusion}
 \label{sec7}
 In this paper, we calculated the $CP$ averaged branching ratios and
 $CP$-violating asymmetries of two-body charmless hadronic $B_{s}$
 ${\to}$ $PP$, $PV$ decays at next-to-leading order in ${\alpha}_{s}$
 with the QCDF approach, including ``nonfactorizable'' corrections, as
 well as those from weak annihilation topologies. We find
 \begin{itemize}
 \item Only several decays, such as $B_{s}$ ${\to}$ $K^{({\ast})}K$,
       $K^{({\ast}){\pm}}{\pi}^{\mp}$, $K^{\pm}{\rho}^{\mp}$,
       ${\eta}^{(\prime)}{\eta}^{(\prime)}$, have large branching ratios,
       which might be accessible at hadron colliders potentially in the near
       future to allow for detailed phenomenological analysis. The $a_{1}$
       dominant decays $B_{s}$ ${\to}$ $K^{({\ast}){\pm}}{\pi}^{\mp}$,
       $K^{\pm}{\rho}^{\mp}$ are generally insensitive to the
       contributions from both ``nonfactorizable'' effects and weak
       annihilation, so the numerical predictions of the QCDF approach on
       their $CP$ averaged branching ratios are similar to their NF's
       counterparts; the main uncertainties come mainly from the CKM
       matrix parameters and form factors. And their direct $CP$-violating
       asymmetries $A_{CP}$ are only a few percent because of their small
       strong phases. But for the branching ratios of $B_{s}$ ${\to}$
       $K^{({\ast})}K$, ${\eta}^{(\prime)}{\eta}^{(\prime)}$ decays, the
       contributions of the ``nonfactorizable'' effects and weak
       annihilation can be sizeable.
 \item Large direct $CP$-violating asymmetries occur in some decays whose
       decay amplitudes are related to coefficient $a_{2}$, because $a_{2}$
       obtains a large imaginary part from ``nonfactorizable'' effects,
       and some can even reach $80\%$ or so, for example,
       $a_{{\epsilon}^{\prime}}^{f}(K_{S}^{0}{\eta})$ ${\approx}$ $-85\%$
       in Table \ref{tab6}. But ${\cal A}_{CP}$ for case-II $CP$ decays is
       very small because of large $x_{s}$. Of course, it is assumed that
       the contribution of power corrections in $1/m_{b}$ are as important
       as radiative ones to strong phase numerically, so the quantitative
       predictions on $CP$-violating asymmetries of $B_{s}$ decays with
       the QCDF approach should not be taken too seriously.
 \item Although we can overconstrain the penguin-to-tree ratio,
       ${\vert}P_{{\pi}{\pi}}/T_{{\pi}{\pi}}{\vert}$, and give a bound
       on ${\gamma}$ from $B_{s}$ decay into charged and neutral kaons,
       too many input parameters bring large theoretical uncertainties,
       what's more, experimental studies on $B_{s}$ decays are very
       limited so far,  so it is not the time to extract some useful
       information on the angle ${\gamma}$ from two-body charmless $B_{s}$
       decays. We look forward to future measurements and theoretical
       developments to give some insight into these parameters.
 \end{itemize}

 \section*{Acknowledgments}
 This work is supported in part by National Natural Science Foundation of
 China. G. Zhu thanks JSPS of Japan for financial support.

 \begin{appendix}

 \section{The decay amplitudes for ${\overline{B}}_{s}^{0}{\to}PP$}
 \label{sec:app1}
 \begin{eqnarray}
 {\cal A}^{f}({\overline{B}}_{s}^{0}{\to}K^{0}{\overline{K}}^{0})
  &=& -i \frac{G_{F}}{\sqrt{2}} f_{K} F_{0}^{B_{s}^{0}{\to}K}
         \Big( m_{B_{s}}^{2}-m_{K}^{2} \Big)
         \Big( V_{ub}V_{us}^{\ast} + V_{cb}V_{cs}^{\ast} \Big)
  \nonumber \\
  & & {\times} \bigg\{ a_{4} - \frac{1}{2}a_{10} + R_{1}
      \Big( a_{6} - \frac{1}{2}a_{8} \Big) \bigg\},
 \label{eq:app1-01}
 \end{eqnarray}
 where
 \begin{math}
  R_{1}= \frac{2m_{{\overline{K}}^{0}}^{2}}
              {(m_{s}+m_{d})(m_{b}-m_{d})}.
 \end{math}
 \begin{eqnarray}
 {\cal A}^{f}({\overline{B}}_{s}^{0}{\to}K^{0}{\pi}^{0})
  &=& -i \frac{G_{F}}{2} f_{\pi} F_{0}^{B_{s}^{0}{\to}K}
         \Big( m_{B_{s}}^{2}-m_{K}^{2} \Big)
         \bigg\{ V_{ub}V_{ud}^{\ast} a_{2}
  \nonumber \\
  & & +  \Big( V_{ub}V_{ud}^{\ast} + V_{cb}V_{cd}^{\ast} \Big)
         \Big[ - a_{4} + \frac{1}{2}a_{10} - \frac{3}{2}
         \Big( a_{7} - a_{9} \Big)
  \nonumber \\
  & & -  R_{2} \Big( a_{6} - \frac{1}{2} a_{8} \Big) \Big] \bigg\},
 \label{eq:app1-02}
 \end{eqnarray}
 where
 \begin{math}
  R_{2}= \frac{2m_{{\pi}^{0}}^{2}}
              {(m_{u}+m_{d})(m_{b}-m_{d})}.
 \end{math}
 \begin{eqnarray}
 {\cal A}^{f}({\overline{B}}_{s}^{0}{\to}K^{0}{\eta}^{(\prime)})
  &=& -i \frac{G_{F}}{\sqrt{2}} f_{K}
         F_{0}^{B_{s}^{0}{\to}{\eta}^{(\prime)}}
         \Big( m_{B_{s}}^{2}-m_{{\eta}^{(\prime)}}^{2} \Big)
         \Big( V_{ub}V_{ud}^{\ast} + V_{cb}V_{cd}^{\ast} \Big)
  \nonumber \\
  & &  {\times} \bigg\{ a_{4} - \frac{1}{2}a_{10}
      +  R_{3} \Big( a_{6} - \frac{1}{2}a_{8} \Big) \bigg\}
  \nonumber \\
  & & \!\!\!\!\!\!\! \!\!\!\!\!\!\! \!\!\!\!\!\!\!
      -i \frac{G_{F}}{\sqrt{2}} f_{{\eta}^{(\prime)}}^{u}
         F_{0}^{B_{s}^{0}{\to}K}
         \Big( m_{B_{s}}^{2}-m_{K}^{2} \Big)
         \bigg\{ V_{ub}V_{ud}^{\ast} a_{2}
  \nonumber \\
  & & \!\!\!\!\!\!\! \!\!\!\!\!\!\! \!\!\!\!\!\!\!
      +  \Big( V_{ub}V_{ud}^{\ast} + V_{cb}V_{cd}^{\ast} \Big)
         \Big[ 2 \Big( a_{3} - a_{5} \Big)
      +  a_{4} - \frac{1}{2}a_{10}
      -  \frac{1}{2} \Big( a_{7} - a_{9} \Big)
  \nonumber \\
  & & \!\!\!\!\!\!\! \!\!\!\!\!\!\! \!\!\!\!\!\!\!
      +  R_{4}^{(\prime)} \Big( a_{6} - \frac{1}{2} a_{8} \Big)
         \Big( 1 - \frac{f_{{\eta}^{(\prime)}}^{u}}
                        {f_{{\eta}^{(\prime)}}^{s}} \Big)
      +  \Big( a_{3} - a_{5} + \frac{1}{2} a_{7}
      -  \frac{1}{2} a_{9} \Big)
         \frac{f_{{\eta}^{(\prime)}}^{s}}{f_{{\eta}^{(\prime)}}^{u}}
         \Big] \bigg\},
 \label{eq:app1-03}
 \end{eqnarray}
 where
 \begin{math}
  R_{3}= \frac{2m_{K^{0}}^{2}}
              {(m_{s}+m_{d})(m_{b}-m_{s})},
 \end{math}
 and
 \begin{math}
  R_{4}^{(\prime)} = \frac{2m_{{\eta}^{(\prime)}}^{2}}
                          {(m_{s}+m_{s})(m_{b}-m_{d})}.
 \end{math}
 \begin{eqnarray}
 {\cal A}^{f}({\overline{B}}_{s}^{0}{\to}{\pi}^{0}{\eta}^{(\prime)})
  &=& -i \frac{G_{F}}{2} f_{\pi}
         F_{0}^{B_{s}^{0}{\to}{\eta}^{(\prime)}}
         \Big( m_{B_{s}}^{2}-m_{{\eta}^{(\prime)}}^{2} \Big)
         \bigg\{ V_{ub}V_{us}^{\ast} a_{2}
  \nonumber \\
  & & +  \Big( V_{ub}V_{us}^{\ast} + V_{cb}V_{cs}^{\ast} \Big)
         \Big[ - \frac{3}{2} \Big( a_{7} - a_{9} \Big) \Big] \bigg\}.
 \label{eq:app1-04}
 \end{eqnarray}
 \begin{eqnarray}
 {\cal A}^{f}({\overline{B}}_{s}^{0}{\to}{\eta}^{(\prime)}{\eta}^{(\prime)})
  &=& -i \sqrt{2} G_{F} f_{{\eta}^{(\prime)}}^{u}
         F_{0}^{B_{s}^{0}{\to}{\eta}^{(\prime)}}
         \Big( m_{B_{s}}^{2}-m_{{\eta}^{(\prime)}}^{2} \Big)
         \bigg\{ V_{ub}V_{us}^{\ast} a_{2}
  \nonumber \\
  & & \!\!\!\!\!\!\! \!\!\!\!\!\!\! \!\!\!\!\!\!\!
      +  \Big( V_{ub}V_{us}^{\ast} + V_{cb}V_{cs}^{\ast} \Big)
         \Big[ 2 \Big( a_{3} - a_{5} \Big)
      -  \frac{1}{2} \Big( a_{7} - a_{9} \Big) \Big]
  \nonumber \\
  & & \!\!\!\!\!\!\! \!\!\!\!\!\!\! \!\!\!\!\!\!\!
      +  \Big( V_{ub}V_{us}^{\ast} + V_{cb}V_{cs}^{\ast} \Big)
         \frac{f_{{\eta}^{(\prime)}}^{s}}{f_{{\eta}^{(\prime)}}^{u}}
         \Big[ a_{3} - a_{5} + a_{4} - \frac{1}{2} a_{10}
      +  \frac{1}{2} a_{7}
  \nonumber \\
  & & \!\!\!\!\!\!\! \!\!\!\!\!\!\! \!\!\!\!\!\!\!
      -  \frac{1}{2} a_{9}
      +  R_{5}^{(\prime)} \Big( a_{6} - \frac{1}{2} a_{8} \Big)
         \Big( 1 - \frac{f_{{\eta}^{(\prime)}}^{u}}
                        {f_{{\eta}^{(\prime)}}^{s}} \Big)
         \Big] \bigg\},
 \label{eq:app1-05}
 \end{eqnarray}
 where
 \begin{math}
  R_{5}^{(\prime)} = \frac{2m_{{\eta}^{(\prime)}}^{2}}
                          {(m_{s}+m_{s})(m_{b}-m_{s})}.
 \end{math}
 \begin{eqnarray}
 {\cal A}^{f}({\overline{B}}_{s}^{0}{\to}{\eta}{\eta}^{\prime})
  &=& -i \frac{G_{F}}{\sqrt{2}} f_{\eta}^{u}
         F_{0}^{B_{s}^{0}{\to}{\eta}^{\prime}}
         \Big( m_{B_{s}}^{2}-m_{{\eta}^{\prime}}^{2} \Big)
         \bigg\{ V_{ub}V_{us}^{\ast} a_{2}
  \nonumber \\
  & & +  \Big( V_{ub}V_{us}^{\ast} + V_{cb}V_{cs}^{\ast} \Big)
         \Big[ 2 \Big( a_{3} - a_{5} \Big)
      -  \frac{1}{2} \Big( a_{7} - a_{9} \Big) \Big]
  \nonumber \\
  & & +  \Big( V_{ub}V_{us}^{\ast} + V_{cb}V_{cs}^{\ast} \Big)
         \frac{f_{\eta}^{s}}{f_{\eta}^{u}}
         \Big[ a_{3} - a_{5} + a_{4} - \frac{1}{2} a_{10}
      +  \frac{1}{2} a_{7}
  \nonumber \\
  & & -  \frac{1}{2} a_{9}
      +  R_{5} \Big( a_{6} - \frac{1}{2} a_{8} \Big)
         \Big( 1 - \frac{f_{\eta}^{u}}{f_{\eta}^{s}} \Big)
         \Big] \bigg\}
  \nonumber \\
  & & -i \frac{G_{F}}{\sqrt{2}} f_{{\eta}^{\prime}}^{u}
         F_{0}^{B_{s}^{0}{\to}{\eta}}
         \Big( m_{B_{s}}^{2}-m_{\eta}^{2} \Big)
         \bigg\{ V_{ub}V_{us}^{\ast} a_{2}
  \nonumber \\
  & & +  \Big( V_{ub}V_{us}^{\ast} + V_{cb}V_{cs}^{\ast} \Big)
         \Big[ 2 \Big( a_{3} - a_{5} \Big)
      -  \frac{1}{2} \Big( a_{7} - a_{9} \Big) \Big]
  \nonumber \\
  & & +  \Big( V_{ub}V_{us}^{\ast} + V_{cb}V_{cs}^{\ast} \Big)
         \frac{f_{{\eta}^{\prime}}^{s}}{f_{{\eta}^{\prime}}^{u}}
         \Big[ a_{3} - a_{5} + a_{4} - \frac{1}{2} a_{10}
      +  \frac{1}{2} a_{7}
  \nonumber \\
  & & -  \frac{1}{2} a_{9}
      +  R_{5}^{\prime} \Big( a_{6} - \frac{1}{2} a_{8} \Big)
         \Big( 1 - \frac{f_{{\eta}^{\prime}}^{u}}
                        {f_{{\eta}^{\prime}}^{s}} \Big)
         \Big] \bigg\}.
 \label{eq:app1-06}
 \end{eqnarray}
 \begin{eqnarray}
 {\cal A}^{f}({\overline{B}}_{s}^{0}{\to}{\pi}^{-}K^{+})
  &=& -i \frac{G_{F}}{\sqrt{2}} f_{\pi} F_{0}^{B_{s}^{0}{\to}K}
         \Big( m_{B_{s}}^{2}-m_{K}^{2} \Big)
         \bigg\{ V_{ub}V_{ud}^{\ast} a_{1}
  \nonumber \\
  & & \!\!\!\!\!\!\! \!\!\!\!\!\!\! \!\!\!\!\!\!\!
      +  \Big( V_{ub}V_{ud}^{\ast} + V_{cb}V_{cd}^{\ast} \Big)
         \Big[ a_{4} + a_{10} + R_{6} \Big( a_{6} + a_{8}
         \Big) \Big] \bigg\},
 \label{eq:app1-07}
 \end{eqnarray}
 where
 \begin{math}
 R_{6} = \frac{2m_{{\pi}^{+}}^{2}}
              {(m_{u}+m_{d})(m_{b}-m_{u})}.
 \end{math}
 \begin{eqnarray}
 {\cal A}^{f}({\overline{B}}_{s}^{0}{\to}K^{-}K^{+})
  &=& -i \frac{G_{F}}{\sqrt{2}} f_{K} F_{0}^{B_{s}^{0}{\to}K}
         \Big( m_{B_{s}}^{2}-m_{K}^{2} \Big)
         \bigg\{ V_{ub}V_{us}^{\ast} a_{1}
  \nonumber \\
  & & \!\!\!\!\!\!\! \!\!\!\!\!\!\! \!\!\!\!\!\!\!
      +  \Big( V_{ub}V_{us}^{\ast} + V_{cb}V_{cs}^{\ast} \Big)
         \Big[ a_{4} + a_{10} + R_{7} \Big( a_{6} + a_{8}
         \Big) \Big] \bigg\},
 \label{eq:app1-08}
 \end{eqnarray}
 where
 \begin{math}
 R_{7} = \frac{2m_{K^{-}}^{2}}
              {(m_{u}+m_{s})(m_{b}-m_{u})}.
 \end{math}

 \section{The decay amplitudes for ${\overline{B}}_{s}^{0}{\to}PV$}
 \label{sec:app2}
 \begin{eqnarray}
 {\cal A}^{f}({\overline{B}}_{s}^{0}{\to}K^{0}{\overline{K}}^{{\ast}0})
  &=&    \sqrt{2} G_{F} f_{K^{\ast}} F_{1}^{B_{s}^{0}{\to}K}
          m_{K^{\ast}} \big( {\epsilon}{\cdot}p_{K^{0}} \big)
         \Big( V_{ub}V_{us}^{\ast} + V_{cb}V_{cs}^{\ast} \Big)
         \bigg\{ a_{4} - \frac{1}{2} a_{10} \bigg\},
 \label{eq:app2-01}
 \end{eqnarray}
 \begin{eqnarray}
 {\cal A}^{f}({\overline{B}}_{s}^{0}{\to}{\overline{K}}^{0}K^{{\ast}0})
  &=&    \sqrt{2} G_{F} f_{K} A_{0}^{B_{s}^{0}{\to}K^{\ast}}
          m_{K^{\ast}} \big( {\epsilon}{\cdot}p_{{\overline{K}}^{0}} \big)
         \Big( V_{ub}V_{us}^{\ast} + V_{cb}V_{cs}^{\ast} \Big)
  \nonumber \\
  & & {\times} \bigg\{ a_{4} - \frac{1}{2} a_{10}
      - Q_{1} \Big( a_{6} - \frac{1}{2} a_{8} \Big) \bigg\},
 \label{eq:app2-02}
 \end{eqnarray}
 where
 \begin{math}
  Q_{1}= \frac{2m_{{\overline{K}}^{0}}^{2}}
              {(m_{s}+m_{d})(m_{b}+m_{d})}.
 \end{math}
 \begin{eqnarray}
 {\cal A}^{f}({\overline{B}}_{s}^{0}{\to}K^{0}{\rho}^{0})
  &=&   G_{F} f_{\rho} F_{1}^{B_{s}^{0}{\to}K}
        m_{\rho} \big( {\epsilon}{\cdot}p_{K^{0}} \big)
        \bigg\{ V_{ub}V_{ud}^{\ast} a_{2}
  \nonumber \\
  & & \!\!\!\!\!\!\! \!\!\!\!\!\!\! \!\!\!\!\!\!\!
      + \Big( V_{ub}V_{ud}^{\ast} + V_{cb}V_{cd}^{\ast} \Big)
        \Big[ - a_{4} + \frac{1}{2} a_{10} + \frac{3}{2} a_{7}
      + \frac{3}{2} a_{9} \Big] \bigg\}.
 \label{eq:app2-03}
 \end{eqnarray}
 \begin{eqnarray}
 {\cal A}^{f}({\overline{B}}_{s}^{0}{\to}K^{0}{\omega})
  &=&   G_{F} f_{\omega} F_{1}^{B_{s}^{0}{\to}K}
        m_{\omega} \big( {\epsilon}{\cdot}p_{K^{0}} \big)
        \bigg\{ V_{ub}V_{ud}^{\ast} a_{2}
  \nonumber \\
  & & \!\!\!\!\!\!\! \!\!\!\!\!\!\! \!\!\!\!\!\!\!
      + \Big( V_{ub}V_{ud}^{\ast} + V_{cb}V_{cd}^{\ast} \Big)
        \Big[ 2 \Big( a_{3} + a_{5} \Big) + a_{4}
      - \frac{1}{2} a_{10} + \frac{1}{2} a_{7}
      + \frac{1}{2} a_{9} \Big] \bigg\}.
 \label{eq:app2-04}
 \end{eqnarray}
 \begin{eqnarray}
 {\cal A}^{f}({\overline{B}}_{s}^{0}{\to}K^{0}{\phi})
  &=&   \sqrt{2} G_{F} m_{\phi}
        \big( {\epsilon}{\cdot}p_{K^{0}} \big)
        \Big( V_{ub}V_{ud}^{\ast} + V_{cb}V_{cd}^{\ast} \Big)
  \nonumber \\
  & & {\times} \bigg\{ f_{\phi} F_{1}^{B_{s}^{0}{\to}K} \Big[ a_{3}
      + a_{5} - \frac{1}{2} a_{7} - \frac{1}{2} a_{9} \Big]
  \nonumber \\
  & & + f_{K} A_{0}^{B_{s}^{0}{\to}{\phi}} \Big[ a_{4}
      - \frac{1}{2} a_{10} - Q_{2} \Big( a_{6}
      - \frac{1}{2} a_{8} \Big) \Big] \bigg\},
 \label{eq:app2-05}
 \end{eqnarray}
 where
 \begin{math}
  Q_{2}= \frac{2m_{K^{0}}^{2}}
              {(m_{s}+m_{d})(m_{b}+m_{s})}.
 \end{math}
 \begin{eqnarray}
 {\cal A}^{f}({\overline{B}}_{s}^{0}{\to}{\pi}^{0}K^{{\ast}0})
  &=&   G_{F} f_{\pi} A_{0}^{B_{s}^{0}{\to}K^{{\ast}0}}
        m_{K^{{\ast}0}} \big( {\epsilon}{\cdot}p_{{\pi}^{0}} \big)
        \bigg\{ V_{ub}V_{ud}^{\ast} a_{2}
  \nonumber \\
  & & + \Big( V_{ub}V_{ud}^{\ast} + V_{cb}V_{cd}^{\ast} \Big)
        \Big[ - a_{4} + \frac{1}{2} a_{10}
      - \frac{3}{2} a_{7}
  \nonumber \\
  & & + \frac{3}{2} a_{9}
      + Q_{3} \Big( a_{6} - \frac{1}{2} a_{8} \Big)
        \Big] \bigg\},
 \label{eq:app2-06}
 \end{eqnarray}
 where
 \begin{math}
  Q_{3}= \frac{2m_{{\pi}^{0}}^{2}}
              {(m_{u}+m_{d})(m_{b}+m_{d})}.
 \end{math}
 \begin{eqnarray}
 {\cal A}^{f}({\overline{B}}_{s}^{0}{\to}{\pi}^{0}{\phi})
  &=&   G_{F} f_{\pi} A_{0}^{B_{s}^{0}{\to}{\phi}}
        m_{\phi} \big( {\epsilon}{\cdot}p_{{\pi}^{0}} \big)
        \bigg\{ V_{ub}V_{us}^{\ast} a_{2}
  \nonumber \\
  & & + \Big( V_{ub}V_{us}^{\ast} + V_{cb}V_{cs}^{\ast} \Big)
        \Big[ - \frac{3}{2} a_{7} + \frac{3}{2} a_{9}
        \Big] \bigg\},
 \label{eq:app2-07}
 \end{eqnarray}
 \begin{eqnarray}
 {\cal A}^{f}({\overline{B}}_{s}^{0}{\to}{\eta}^{(\prime)}K^{{\ast}0})
  &=&   \sqrt{2} G_{F} m_{K^{{\ast}0}}
        \big( {\epsilon}{\cdot}p_{{\eta}^{(\prime)}} \big)
        \bigg\{ f_{{\eta}^{(\prime)}}^{u}
        A_{0}^{B_{s}^{0}{\to}K^{{\ast}0}}
        \Big[ V_{ub}V_{ud}^{\ast} a_{2}
  \nonumber \\
  & & \!\!\!\!\!\!\! \!\!\!\!\!\!\! \!\!\!\!\!\!\!
      + \Big( V_{ub}V_{ud}^{\ast} + V_{cb}V_{cd}^{\ast} \Big)
        \Big( 2 \big( a_{3} - a_{5} \big) + a_{4}
      - \frac{1}{2} a_{10} - \frac{1}{2} a_{7}
  \nonumber \\
  & & 
      + \frac{1}{2} a_{9}
      - Q_{4}^{(\prime)} \big( a_{6} - \frac{1}{2} a_{8} \big)
        \big( 1 - \frac{f_{{\eta}^{(\prime)}}^{u}}
             {f_{{\eta}^{(\prime)}}^{s}} \big) \Big)
  \nonumber \\
  & & \!\!\!\!\!\!\! \!\!\!\!\!\!\! \!\!\!\!\!\!\!
      + \Big( V_{ub}V_{ud}^{\ast} + V_{cb}V_{cd}^{\ast} \Big)
        \frac{f_{{\eta}^{(\prime)}}^{s}}{f_{{\eta}^{(\prime)}}^{u}}
        \Big( a_{3} - a_{5} + \frac{1}{2} a_{7} - \frac{1}{2} a_{9}
        \Big) \Big]
  \nonumber \\
  & & \!\!\!\!\!\!\! \!\!\!\!\!\!\! \!\!\!\!\!\!\!
      + f_{K^{\ast}} F_{1}^{B_{s}^{0}{\to}{\eta}^{(\prime)}}
        \Big( V_{ub}V_{ud}^{\ast} + V_{cb}V_{cd}^{\ast} \Big)
        \Big( a_{4} - \frac{1}{2} a_{10} \Big) \bigg\},
 \label{eq:app2-08}
 \end{eqnarray}
 where
 \begin{math}
  Q_{4}^{(\prime)} = \frac{2m_{{\eta}^{(\prime)}}^{2}}
              {(m_{s}+m_{s})(m_{b}+m_{s})}.
 \end{math}
 \begin{eqnarray}
 {\cal A}^{f}({\overline{B}}_{s}^{0}{\to}{\eta}^{(\prime)}{\rho}^{0})
  &=&   G_{F} f_{\rho} F_{1}^{B_{s}^{0}{\to}{\eta}^{(\prime)}}
        m_{\rho} \big( {\epsilon}{\cdot}p_{{\eta}^{(\prime)}} \big)
        \bigg\{ V_{ub}V_{us}^{\ast} a_{2}
  \nonumber \\
  & & + \Big( V_{ub}V_{us}^{\ast} + V_{cb}V_{cs}^{\ast} \Big)
        \Big[ \frac{3}{2} \Big( a_{7} + a_{9} \Big) \Big] \bigg\},
 \label{eq:app2-09}
 \end{eqnarray}
 \begin{eqnarray}
 {\cal A}^{f}({\overline{B}}_{s}^{0}{\to}{\eta}^{(\prime)}{\omega})
  &=&   G_{F} f_{\omega} F_{1}^{B_{s}^{0}{\to}{\eta}^{(\prime)}}
        m_{\omega} \big( {\epsilon}{\cdot}p_{{\eta}^{(\prime)}} \big)
        \bigg\{ V_{ub}V_{us}^{\ast} a_{2}
  \nonumber \\
  & & \!\!\!\!\!\!\! \!\!\!\!\!\!\! \!\!\!\!\!\!\!
      + \Big( V_{ub}V_{us}^{\ast} + V_{cb}V_{cs}^{\ast} \Big)
        \Big[ 2 \Big( a_{3} + a_{5} \Big)
      + \frac{1}{2} \Big( a_{7} + a_{9} \Big) \Big] \bigg\},
 \label{eq:app2-10}
 \end{eqnarray}
 \begin{eqnarray}
 {\cal A}^{f}({\overline{B}}_{s}^{0}{\to}{\eta}^{(\prime)}{\phi})
  &=&   \sqrt{2} G_{F} m_{\phi}
        \big( {\epsilon}{\cdot}p_{{\eta}^{(\prime)}} \big)
        \bigg\{ f_{{\eta}^{(\prime)}}^{u} A_{0}^{B_{s}^{0}{\to}{\phi}}
        \Big[ V_{ub}V_{us}^{\ast} a_{2}
  \nonumber \\
  & & \!\!\!\!\!\!\! \!\!\!\!\!\!\! \!\!\!\!\!\!\!
      + \Big( V_{ub}V_{us}^{\ast} + V_{cb}V_{cs}^{\ast} \Big)
        \Big( 2 \big( a_{3} - a_{5} \big)
      - \frac{1}{2} \big( a_{7} - a_{9} \big) \Big)
  \nonumber \\
   & & \!\!\!\!\!\!\! \!\!\!\!\!\!\! \!\!\!\!\!\!\!
      + \Big( V_{ub}V_{us}^{\ast} + V_{cb}V_{cs}^{\ast} \Big)
        \frac{f_{{\eta}^{(\prime)}}^{s}}{f_{{\eta}^{(\prime)}}^{u}}
        \Big( a_{3} - a_{5} + a_{4} - \frac{1}{2} a_{10}
   \nonumber \\
   & & \!\!\!\!\!\!\! \!\!\!\!\!\!\! \!\!\!\!\!\!\!
      + \frac{1}{2} a_{7} - \frac{1}{2} a_{9}
      - Q_{4}^{(\prime)} \big( a_{6} - \frac{1}{2} a_{8} \big)
        \big( 1 - \frac{f_{{\eta}^{(\prime)}}^{u}}
        {f_{{\eta}^{(\prime)}}^{s}} \big) \Big) \Big]
   \nonumber \\
   & & \!\!\!\!\!\!\! \!\!\!\!\!\!\! \!\!\!\!\!\!\!
      + f_{\phi} F_{1}^{B_{s}^{0}{\to}{\eta}^{(\prime)}}
        \Big( V_{ub}V_{us}^{\ast} + V_{cb}V_{cs}^{\ast} \Big)
        \Big( a_{3} + a_{5}
   \nonumber \\
   & & 
      + a_{4} - \frac{1}{2} a_{10} - \frac{1}{2} a_{7}
      - \frac{1}{2} a_{9} \Big) \bigg\},
 \label{eq:app2-11}
 \end{eqnarray}
 \begin{eqnarray}
 {\cal A}^{f}({\overline{B}}_{s}^{0}{\to}K^{+}{\rho}^{-})
  &=&   \sqrt{2} G_{F} f_{\rho} F_{1}^{B_{s}^{0}{\to}K}
        m_{\rho} \big( {\epsilon}{\cdot}p_{K} \big)
        \bigg\{ V_{ub}V_{ud}^{\ast} a_{1}
  \nonumber \\
  & & 
      + \Big( V_{ub}V_{ud}^{\ast} + V_{cb}V_{cd}^{\ast} \Big)
        \Big( a_{4} + a_{10} \Big) \bigg\},
 \label{eq:app2-12}
 \end{eqnarray}
 \begin{eqnarray}
 {\cal A}^{f}({\overline{B}}_{s}^{0}{\to}K^{+}K^{{\ast}-})
  &=&   \sqrt{2} G_{F} f_{K^{\ast}} F_{1}^{B_{s}^{0}{\to}K}
        m_{K^{\ast}} \big( {\epsilon}{\cdot}p_{K} \big)
        \bigg\{ V_{ub}V_{us}^{\ast} a_{1}
  \nonumber \\
  & & + \Big( V_{ub}V_{us}^{\ast} + V_{cb}V_{cs}^{\ast} \Big)
        \Big( a_{4} + a_{10} \Big) \bigg\},
 \label{eq:app2-13}
 \end{eqnarray}
 \begin{eqnarray}
 {\cal A}^{f}({\overline{B}}_{s}^{0}{\to}K^{-}K^{{\ast}+})
  &=&   \sqrt{2} G_{F} f_{K} A_{0}^{B_{s}^{0}{\to}K^{\ast}}
        m_{K^{\ast}} \big( {\epsilon}{\cdot}p_{K} \big)
        \bigg\{ V_{ub}V_{us}^{\ast} a_{1}
  \nonumber \\
  & & \!\!\!\!\!\!\! \!\!\!\!\!\!\! \!\!\!\!\!\!\!
      + \Big( V_{ub}V_{us}^{\ast} + V_{cb}V_{cs}^{\ast} \Big)
        \Big[ a_{4} + a_{10} - Q_{5} \Big( a_{6}
      + a_{8} \Big) \Big] \bigg\},
 \label{eq:app2-14}
 \end{eqnarray}
 where
 \begin{math}
  Q_{5} = \frac{2m_{K^{-}}^{2}}
               {(m_{u}+m_{s})(m_{b}+m_{u})}.
 \end{math}
 \begin{eqnarray}
 {\cal A}^{f}({\overline{B}}_{s}^{0}{\to}{\pi}^{-}K^{{\ast}+})
  &=&   \sqrt{2} G_{F} f_{\pi} A_{0}^{B_{s}^{0}{\to}K^{\ast}}
        m_{K^{\ast}} \big( {\epsilon}{\cdot}p_{\pi} \big)
        \bigg\{ V_{ub}V_{ud}^{\ast} a_{1}
  \nonumber \\
  & & \!\!\!\!\!\!\! \!\!\!\!\!\!\! \!\!\!\!\!\!\!
      + \Big( V_{ub}V_{ud}^{\ast} + V_{cb}V_{cd}^{\ast} \Big)
        \Big[ a_{4} + a_{10} - Q_{6} \Big( a_{6}
      + a_{8} \Big) \Big] \bigg\},
 \label{eq:app2-15}
 \end{eqnarray}
 where
 \begin{math}
  Q_{6} = \frac{2m_{{\pi}^{-}}^{2}}
               {(m_{u}+m_{d})(m_{b}+m_{u})}.
 \end{math}

 \section{The weak annihilation amplitudes
          for ${\overline{B}}_{s}^{0}{\to}PP$}
 \label{sec:app3}
 \begin{eqnarray}
 {\cal A}^{a}({\overline{B}}_{s}^{0}{\to}K^{0}{\overline{K}}^{0})
  &=& -i \frac{G_{F}}{\sqrt{2}} f_{B_{s}} f_{K}^{2} \bigg\{
         \Big( V_{ub}V_{us}^{\ast} + V_{cb}V_{cs}^{\ast} \Big)
         \Big[ b_{3}({\overline{K}}^{0},K^{0})
      +  b_{4}({\overline{K}}^{0},K^{0})
  \nonumber \\
  & & \!\!\!\!\!\!\! \!\!\!\!\!\!\! \!\!\!\!\!\!\!
      +  b_{4}(K^{0},{\overline{K}}^{0})
      - \frac{1}{2} b_{3}^{ew}({\overline{K}}^{0},K^{0})
      - \frac{1}{2} b_{4}^{ew}({\overline{K}}^{0},K^{0})
      - \frac{1}{2} b_{4}^{ew}(K^{0},{\overline{K}}^{0})
        \Big] \bigg\},
 \label{eq:app3-01}
 \end{eqnarray}
 \begin{eqnarray}
 {\cal A}^{a}({\overline{B}}_{s}^{0}{\to}K^{0}{\pi}^{0})
  &=& i \frac{G_{F}}{2} f_{B_{s}} f_{K} f_{\pi}
        \Big( V_{ub}V_{ud}^{\ast} + V_{cb}V_{cd}^{\ast} \Big)
        \bigg\{ b_{3}({\pi}^{0},K^{0})
      - \frac{1}{2} b_{3}^{ew}({\pi}^{0},K^{0}) \bigg\},
 \label{eq:app3-02}
 \end{eqnarray}
 \begin{eqnarray}
 {\cal A}^{a}({\overline{B}}_{s}^{0}{\to}K^{0}{\eta}^{(\prime)})
  &=& -i \frac{G_{F}}{\sqrt{2}} f_{B_{s}} f_{K}
         \Big( V_{ub}V_{ud}^{\ast} + V_{cb}V_{cd}^{\ast} \Big)
         \bigg\{ f^{u}_{{\eta}^{(\prime)}}
         \Big[ b_{3}({\eta}^{(\prime)},K^{0})
  \nonumber \\
  & & \!\!\!\!\!\!\! \!\!\!\!\!\!\! \!\!\!\!\!\!\!
      -  \frac{1}{2} b_{3}^{ew}({\eta}^{(\prime)},K^{0}) \Big]
      +  f^{s}_{{\eta}^{(\prime)}} \Big[
         b_{3}(K^{0},{\eta}^{(\prime)})
      -  \frac{1}{2} b_{3}^{ew}(K^{0},{\eta}^{(\prime)})
         \Big] \bigg\},
 \label{eq:app3-03}
 \end{eqnarray}
 \begin{eqnarray}
 {\cal A}^{a}({\overline{B}}_{s}^{0}{\to}{\pi}^{0}{\pi}^{0})
  &=& -i \frac{G_{F}}{\sqrt{2}} f_{B_{s}} f_{\pi}^{2} \bigg\{
         V_{ub}V_{us}^{\ast} b_{1}({\pi}^{0},{\pi}^{0})
  \nonumber \\
  & & 
      +  \Big( V_{ub}V_{us}^{\ast} + V_{cb}V_{cs}^{\ast} \Big)
         \Big[ 2 b_{4}({\pi}^{0},{\pi}^{0})
      +  \frac{1}{2} b_{4}^{ew}({\pi}^{0},{\pi}^{0})
         \Big] \bigg\},
 \label{eq:app3-04}
 \end{eqnarray}
 \begin{eqnarray}
 {\cal A}^{a}({\overline{B}}_{s}^{0}{\to}{\pi}^{0}{\eta}^{(\prime)})
  &=& -i \frac{G_{F}}{2} f_{B_{s}} f_{\pi} f^{u}_{{\eta}^{(\prime)}}
         \bigg\{ V_{ub}V_{us}^{\ast} \Big[
         b_{1}({\pi}^{0},{\eta}^{(\prime)})
      +  b_{1}({\eta}^{(\prime)},{\pi}^{0}) \Big]
  \nonumber \\
  & & 
      +  \Big( V_{ub}V_{us}^{\ast} + V_{cb}V_{cs}^{\ast} \Big)
         \Big[ \frac{3}{2} b_{4}^{ew}({\pi}^{0},{\eta}^{(\prime)})
      +  \frac{3}{2} b_{4}^{ew}({\eta}^{(\prime)},{\pi}^{0})
         \Big] \bigg\},
 \label{eq:app3-05}
 \end{eqnarray}
 \begin{eqnarray}
 {\cal A}^{a}({\overline{B}}_{s}^{0}{\to}{\eta}^{(\prime)}{\eta}^{(\prime)})
  &=& -i \sqrt{2} G_{F} f_{B_{s}} \bigg\{ f^{u}_{{\eta}^{(\prime)}}
         f^{u}_{{\eta}^{(\prime)}} \Big[ V_{ub}V_{us}^{\ast}
         b_{1}({\eta}^{(\prime)},{\eta}^{(\prime)})
  \nonumber \\
  & & 
      +  \Big( V_{ub}V_{us}^{\ast} + V_{cb}V_{cs}^{\ast} \Big)
         \Big( 2 b_{4}({\eta}^{(\prime)},{\eta}^{(\prime)})
      +  \frac{1}{2} b_{4}^{ew}({\eta}^{(\prime)},{\eta}^{(\prime)})
         \Big) \Big]
  \nonumber \\
  & & 
      +  f^{s}_{{\eta}^{(\prime)}} f^{s}_{{\eta}^{(\prime)}}
         \Big( V_{ub}V_{us}^{\ast} + V_{cb}V_{cs}^{\ast} \Big)
         \Big[ b_{3}({\eta}^{(\prime)},{\eta}^{(\prime)})
      +  b_{4}({\eta}^{(\prime)},{\eta}^{(\prime)})
  \nonumber \\
  & & 
      -  \frac{1}{2} b_{3}^{ew}({\eta}^{(\prime)},{\eta}^{(\prime)})
      -  \frac{1}{2} b_{4}^{ew}({\eta}^{(\prime)},{\eta}^{(\prime)})
         \Big] \bigg\},
 \label{eq:app3-06}
 \end{eqnarray}
 \begin{eqnarray}
 {\cal A}^{a}({\overline{B}}_{s}^{0}{\to}{\eta}{\eta}^{\prime})
  &=& -i \frac{G_{F}}{\sqrt{2}} f_{B_{s}} f^{u}_{\eta}
         f^{u}_{{\eta}^{\prime}} \bigg\{ V_{ub}V_{us}^{\ast}
         \Big[ b_{1}({\eta}^{\prime},{\eta})
      +  b_{1}({\eta},{\eta}^{\prime}) \Big]
  \nonumber \\
  & & 
      +  \Big( V_{ub}V_{us}^{\ast} + V_{cb}V_{cs}^{\ast} \Big)
         \Big[ 2 b_{4}({\eta}^{\prime},{\eta})
      +  2 b_{4}({\eta},{\eta}^{\prime})
  \nonumber \\
  & & 
      +  \frac{1}{2} b_{4}^{ew}({\eta}^{\prime},{\eta})
      +  \frac{1}{2} b_{4}^{ew}({\eta},{\eta}^{\prime})
         \Big] \bigg\}
  \nonumber \\
  & & 
      -i \frac{G_{F}}{\sqrt{2}} f_{B_{s}} f^{s}_{\eta}
         f^{s}_{{\eta}^{\prime}}
         \Big( V_{ub}V_{us}^{\ast} + V_{cb}V_{cs}^{\ast} \Big)
         \bigg\{ b_{3}({\eta}^{\prime},{\eta})
      +  b_{3}({\eta},{\eta}^{\prime})
  \nonumber \\
  & & 
      +  b_{4}({\eta}^{\prime},{\eta})
      +  b_{4}({\eta},{\eta}^{\prime})
      -  \frac{1}{2} b_{3}^{ew}({\eta}^{\prime},{\eta})
      -  \frac{1}{2} b_{3}^{ew}({\eta},{\eta}^{\prime})
  \nonumber \\
  & & 
      -  \frac{1}{2} b_{4}^{ew}({\eta}^{\prime},{\eta})
      -  \frac{1}{2} b_{4}^{ew}({\eta},{\eta}^{\prime})
         \bigg\},
 \label{eq:app3-07}
 \end{eqnarray}
 \begin{eqnarray}
 {\cal A}^{a}({\overline{B}}_{s}^{0}{\to}{\pi}^{-}K^{+})
  &=& -i \frac{G_{F}}{\sqrt{2}} f_{B_{s}} f_{\pi} f_{K}
         \Big( V_{ub}V_{ud}^{\ast} + V_{cb}V_{cd}^{\ast} \Big)
         \bigg\{ b_{3}({\pi}^{-},K^{+})
      -  \frac{1}{2} b_{3}^{ew}({\pi}^{-},K^{+}) \bigg\},
 \label{eq:app3-08}
 \end{eqnarray}
 \begin{eqnarray}
 {\cal A}^{a}({\overline{B}}_{s}^{0}{\to}K^{-}K^{+})
  &=& -i \frac{G_{F}}{\sqrt{2}} f_{B_{s}} f_{K}^{2} \bigg\{
         V_{ub}V_{us}^{\ast} b_{1}(K^{+},K^{-})
  \nonumber \\
  & & \!\!\!\!\!\!\! \!\!\!\!\!\!\! \!\!\!\!\!\!\!
      +  \Big( V_{ub}V_{us}^{\ast} + V_{cb}V_{cs}^{\ast} \Big)
         \Big[ b_{3}(K^{-},K^{+}) + b_{4}(K^{+},K^{-})
      +  b_{4}(K^{-},K^{+})
  \nonumber \\
  & & \!\!\!\!\!\!\! \!\!\!\!\!\!\! \!\!\!\!\!\!\!
      - \frac{1}{2} b_{3}^{ew}(K^{-},K^{+})
      + b_{4}^{ew}(K^{+},K^{-})
      -  \frac{1}{2} b_{4}^{ew}(K^{-},K^{+}) \Big] \bigg\},
 \label{eq:app3-09}
 \end{eqnarray}
 \begin{eqnarray}
 {\cal A}^{a}({\overline{B}}_{s}^{0}{\to}{\pi}^{-}{\pi}^{+})
  &=& -i \frac{G_{F}}{\sqrt{2}} f_{B_{s}} f_{\pi}^{2} \bigg\{
         V_{ub}V_{us}^{\ast} b_{1}({\pi}^{+},{\pi}^{-})
      +  \Big( V_{ub}V_{us}^{\ast} + V_{cb}V_{cs}^{\ast} \Big)
         \Big[ b_{4}({\pi}^{+},{\pi}^{-})
  \nonumber \\
  & &  
      + b_{4}({\pi}^{-},{\pi}^{+})
      + b_{4}^{ew}({\pi}^{+},{\pi}^{-})
      -  \frac{1}{2} b_{4}^{ew}({\pi}^{-},{\pi}^{+}) \Big] \bigg\},
 \label{eq:app3-10}
 \end{eqnarray}

 \section{The weak annihilation amplitudes
          for ${\overline{B}}_{s}^{0}{\to}PV$}
 \label{sec:app4}
 \begin{eqnarray}
 {\cal A}^{a}({\overline{B}}_{s}^{0}{\to}K^{0}{\rho}^{0})
  &=& - \frac{G_{F}}{2} f_{B_{s}} f_{K} f_{\rho}
        \Big( V_{ub}V_{ud}^{\ast} + V_{cb}V_{cd}^{\ast} \Big)
        \bigg\{ b_{3}({\rho}^{0},K^{0})
      - \frac{1}{2} b_{3}^{ew}({\rho}^{0},K^{0}) \bigg\},
 \label{eq:app4-01}
 \end{eqnarray}
 \begin{eqnarray}
 {\cal A}^{a}({\overline{B}}_{s}^{0}{\to}K^{0}{\omega})
  &=&   \frac{G_{F}}{2} f_{B_{s}} f_{K} f_{\omega}
        \Big( V_{ub}V_{ud}^{\ast} + V_{cb}V_{cd}^{\ast} \Big)
        \bigg\{ b_{3}({\omega},K^{0})
      - \frac{1}{2} b_{3}^{ew}({\omega},K^{0}) \bigg\},
 \label{eq:app4-02}
 \end{eqnarray}
 \begin{eqnarray}
 {\cal A}^{a}({\overline{B}}_{s}^{0}{\to}K^{0}{\phi})
  &=&   \frac{G_{F}}{\sqrt{2}} f_{B_{s}} f_{K} f_{\phi}
        \Big( V_{ub}V_{ud}^{\ast} + V_{cb}V_{cd}^{\ast} \Big)
        \bigg\{ b_{3}(K^{0},{\phi})
      - \frac{1}{2} b_{3}^{ew}(K^{0},{\phi}) \bigg\},
 \label{eq:app4-03}
 \end{eqnarray}
 \begin{eqnarray}
 {\cal A}^{a}({\overline{B}}_{s}^{0}{\to}K^{0}{\overline{K}}^{{\ast}0})
  &=&   \frac{G_{F}}{\sqrt{2}} f_{B_{s}} f_{K} f_{K^{\ast}}
        \Big( V_{ub}V_{us}^{\ast} + V_{cb}V_{cs}^{\ast} \Big)
        \bigg\{ b_{3}({\overline{K}}^{{\ast}0},K^{0})
      + b_{4}(K^{0},{\overline{K}}^{{\ast}0})
  \nonumber \\
  & & \!\!\!\!\!\!\! \!\!\!\!\!\!\! \!\!\!\!\!\!\!
      + b_{4}({\overline{K}}^{{\ast}0},K^{0})
      - \frac{1}{2} b_{3}^{ew}({\overline{K}}^{{\ast}0},K^{0})
      - \frac{1}{2} b_{4}^{ew}(K^{0},{\overline{K}}^{{\ast}0})
      - \frac{1}{2} b_{4}^{ew}({\overline{K}}^{{\ast}0},K^{0})
        \bigg\},
 \label{eq:app4-04}
 \end{eqnarray}
 \begin{eqnarray}
 {\cal A}^{a}({\overline{B}}_{s}^{0}{\to}{\overline{K}}^{0}K^{{\ast}0})
  &=&   \frac{G_{F}}{\sqrt{2}} f_{B_{s}} f_{K} f_{K^{\ast}}
        \Big( V_{ub}V_{us}^{\ast} + V_{cb}V_{cs}^{\ast} \Big)
        \bigg\{ b_{3}({\overline{K}}^{0},K^{{\ast}0})
      + b_{4}(K^{{\ast}0},{\overline{K}}^{0})
  \nonumber \\
  & & \!\!\!\!\!\!\! \!\!\!\!\!\!\! \!\!\!\!\!\!\!
      + b_{4}({\overline{K}}^{0},K^{{\ast}0})
      - \frac{1}{2} b_{3}^{ew}({\overline{K}}^{0},K^{{\ast}0})
      - \frac{1}{2} b_{4}^{ew}(K^{{\ast}0},{\overline{K}}^{0})
      - \frac{1}{2} b_{4}^{ew}({\overline{K}}^{0},K^{{\ast}0})
        \bigg\},
 \label{eq:app4-05}
 \end{eqnarray}
 \begin{eqnarray}
 {\cal A}^{a}({\overline{B}}_{s}^{0}{\to}{\pi}^{0}K^{{\ast}0})
  &=& - \frac{G_{F}}{2} f_{B_{s}} f_{\pi} f_{K^{\ast}}
        \Big( V_{ub}V_{ud}^{\ast} + V_{cb}V_{cd}^{\ast} \Big)
        \bigg\{ b_{3}({\pi}^{0},K^{{\ast}0})
      - \frac{1}{2} b_{3}^{ew}({\pi}^{0},K^{{\ast}0}) \bigg\},
 \label{eq:app4-06}
 \end{eqnarray}
 \begin{eqnarray}
 {\cal A}^{a}({\overline{B}}_{s}^{0}{\to}{\pi}^{0}{\rho}^{0})
  &=&   \frac{G_{F}}{2\sqrt{2}} f_{B_{s}} f_{\pi} f_{\rho}
        \bigg\{ V_{ub}V_{us}^{\ast}
        \Big[ b_{1}({\pi}^{0},{\rho}^{0})
      + b_{1}({\rho}^{0},{\pi}^{0}) \Big]
  \nonumber \\
  & & 
      + \Big( V_{ub}V_{us}^{\ast} + V_{cb}V_{cs}^{\ast} \Big)
        \Big[ 2 b_{4}({\pi}^{0},{\rho}^{0})
      + 2 b_{4}({\rho}^{0},{\pi}^{0})
  \nonumber \\
  & & 
      + \frac{1}{2} b_{4}^{ew}({\pi}^{0},{\rho}^{0})
      + \frac{1}{2} b_{4}^{ew}({\rho}^{0},{\pi}^{0})
        \Big] \bigg\},
 \label{eq:app4-07}
 \end{eqnarray}
 \begin{eqnarray}
 {\cal A}^{a}({\overline{B}}_{s}^{0}{\to}{\pi}^{0}{\omega})
  &=&   \frac{G_{F}}{2\sqrt{2}} f_{B_{s}} f_{\pi} f_{\omega}
        \bigg\{ V_{ub}V_{us}^{\ast}
        \Big[ b_{1}({\pi}^{0},{\omega})
      + b_{1}({\omega},{\pi}^{0}) \Big]
  \nonumber \\
  & & 
      + \Big( V_{ub}V_{us}^{\ast} + V_{cb}V_{cs}^{\ast} \Big)
        \Big[ \frac{3}{2} b_{4}^{ew}({\pi}^{0},{\omega})
      + \frac{3}{2} b_{4}^{ew}({\omega},{\pi}^{0}) \Big] \bigg\},
 \label{eq:app4-08}
 \end{eqnarray}
 \begin{equation}
 {\cal A}^{a}({\overline{B}}_{s}^{0}{\to}{\pi}^{0}{\phi})=0,
 \label{eq:app4-09}
 \end{equation}
 \begin{eqnarray}
 {\cal A}^{a}({\overline{B}}_{s}^{0}{\to}{\eta}^{(\prime)}K^{{\ast}0})
  &=&   \frac{G_{F}}{\sqrt{2}} f_{B_{s}} f_{K^{\ast}}
        \Big( V_{ub}V_{ud}^{\ast} + V_{cb}V_{cd}^{\ast} \Big)
        \bigg\{ f^{u}_{{\eta}^{(\prime)}} \Big[
        b_{3}({\eta}^{(\prime)},K^{{\ast}0})
  \nonumber \\
  & & \!\!\!\!\!\!\! \!\!\!\!\!\!\! \!\!\!\!\!\!\!
      - \frac{1}{2} b_{3}^{ew}({\eta}^{(\prime)},K^{{\ast}0}) \Big]
      + f^{s}_{{\eta}^{(\prime)}} \Big[
        b_{3}(K^{{\ast}0},{\eta}^{(\prime)})
      - \frac{1}{2} b_{3}^{ew}(K^{{\ast}0},{\eta}^{(\prime)})
        \Big] \bigg\},
 \label{eq:app4-10}
 \end{eqnarray}
 \begin{eqnarray}
 {\cal A}^{a}({\overline{B}}_{s}^{0}{\to}{\eta}^{(\prime)}{\phi})
  &=&   \frac{G_{F}}{\sqrt{2}} f_{B_{s}} f_{\phi}
        f^{s}_{{\eta}^{(\prime)}}
        \Big( V_{ub}V_{us}^{\ast} + V_{cb}V_{cs}^{\ast} \Big)
        \bigg\{ b_{3}({\eta}^{(\prime)},{\phi})
      + b_{3}({\phi},{\eta}^{(\prime)})
  \nonumber \\
  & & 
      + b_{4}({\eta}^{(\prime)},{\phi})
      + b_{4}({\phi},{\eta}^{(\prime)})
      - \frac{1}{2} b_{3}^{ew}({\eta}^{(\prime)},{\phi})
      - \frac{1}{2} b_{3}^{ew}({\phi},{\eta}^{(\prime)})
  \nonumber \\
  & & 
      - \frac{1}{2} b_{4}^{ew}({\eta}^{(\prime)},{\phi})
      - \frac{1}{2} b_{4}^{ew}({\phi},{\eta}^{(\prime)}) \bigg\},
 \label{eq:app4-11}
 \end{eqnarray}
 \begin{eqnarray}
 {\cal A}^{a}({\overline{B}}_{s}^{0}{\to}{\eta}^{(\prime)}{\rho}^{0})
  &=&   \frac{G_{F}}{2} f_{B_{s}} f_{\rho} f^{u}_{{\eta}^{(\prime)}}
        \bigg\{ V_{ub}V_{us}^{\ast} \Big[
        b_{1}({\eta}^{(\prime)},{\rho}^{0})
      + b_{1}({\rho}^{0},{\eta}^{(\prime)}) \Big]
  \nonumber \\
  & & 
      + \Big( V_{ub}V_{us}^{\ast} + V_{cb}V_{cs}^{\ast} \Big)
        \Big[ \frac{3}{2} b_{4}^{ew}({\eta}^{(\prime)},{\rho}^{0})
      + \frac{3}{2} b_{4}^{ew}({\rho}^{0},{\eta}^{(\prime)})
        \Big] \bigg\},
 \label{eq:app4-12}
 \end{eqnarray}
 \begin{eqnarray}
 {\cal A}^{a}({\overline{B}}_{s}^{0}{\to}{\eta}^{(\prime)}{\omega})
  &=&   \frac{G_{F}}{2} f_{B_{s}} f_{\omega} f^{u}_{{\eta}^{(\prime)}}
        \bigg\{ V_{ub}V_{us}^{\ast} \Big[
        b_{1}({\eta}^{(\prime)},{\omega})
      + b_{1}({\omega},{\eta}^{(\prime)}) \Big]
  \nonumber \\
  & & 
      + \Big( V_{ub}V_{us}^{\ast} + V_{cb}V_{cs}^{\ast} \Big)
        \Big[ 2 b_{4}({\eta}^{(\prime)},{\omega})
      + 2 b_{4}({\omega},{\eta}^{(\prime)})
  \nonumber \\
  & & 
      + \frac{1}{2} b_{4}^{ew}({\eta}^{(\prime)},{\omega})
      + \frac{1}{2} b_{4}^{ew}({\omega},{\eta}^{(\prime)})
        \Big] \bigg\},
 \label{eq:app4-13}
 \end{eqnarray}
 \begin{eqnarray}
 {\cal A}^{a}({\overline{B}}_{s}^{0}{\to}K^{+}{\rho}^{-})
  &=&   \frac{G_{F}}{\sqrt{2}} f_{B_{s}} f_{\rho} f_{K}
        \Big( V_{ub}V_{ud}^{\ast} + V_{cb}V_{cd}^{\ast} \Big)
        \bigg\{ b_{3}({\rho}^{-},K^{+})
      - \frac{1}{2} b_{3}^{ew}({\rho}^{-},K^{+}) \bigg\},
 \label{eq:app4-14}
 \end{eqnarray}
 \begin{eqnarray}
 {\cal A}^{a}({\overline{B}}_{s}^{0}{\to}K^{+}K^{{\ast}-})
  &=&   \frac{G_{F}}{\sqrt{2}} f_{B_{s}} f_{K} f_{K^{\ast}} \bigg\{
        V_{ub}V_{us}^{\ast} \Big[ b_{1}(K^{+},K^{{\ast}-})
  \nonumber \\
  & & \!\!\!\!\!\!\! \!\!\!\!\!\!\! \!\!\!\!\!\!\!
      + \Big( V_{ub}V_{us}^{\ast} + V_{cb}V_{cs}^{\ast} \Big)
        \Big[ b_{3}(K^{{\ast}-},K^{+})
      + b_{4}(K^{+},K^{{\ast}-}) + b_{4}(K^{{\ast}-},K^{+})
  \nonumber \\
  & & \!\!\!\!\!\!\! \!\!\!\!\!\!\! \!\!\!\!\!\!\!
      - \frac{1}{2} b_{3}^{ew}(K^{{\ast}-},K^{+})
      + b_{4}^{ew}(K^{+},K^{{\ast}-})
      - \frac{1}{2} b_{4}^{ew}(K^{{\ast}-},K^{+})
        \Big] \bigg\},
 \label{eq:app4-15}
 \end{eqnarray}
 \begin{eqnarray}
 {\cal A}^{a}({\overline{B}}_{s}^{0}{\to}K^{-}K^{{\ast}+})
  &=&   \frac{G_{F}}{\sqrt{2}} f_{B_{s}} f_{K} f_{K^{\ast}} \bigg\{
        V_{ub}V_{us}^{\ast} \Big[ b_{1}(K^{{\ast}+},K^{-})
  \nonumber \\
  & & \!\!\!\!\!\!\! \!\!\!\!\!\!\! \!\!\!\!\!\!\!
      + \Big( V_{ub}V_{us}^{\ast} + V_{cb}V_{cs}^{\ast} \Big)
        \Big[ b_{3}(K^{-},K^{{\ast}+})
      + b_{4}(K^{{\ast}+},K^{-}) + b_{4}(K^{-},K^{{\ast}+})
  \nonumber \\
  & & \!\!\!\!\!\!\! \!\!\!\!\!\!\! \!\!\!\!\!\!\!
      - \frac{1}{2} b_{3}^{ew}(K^{-},K^{{\ast}+})
      + b_{4}^{ew}(K^{{\ast}+},K^{-})
      - \frac{1}{2} b_{4}^{ew}(K^{-},K^{{\ast}+})
        \Big] \bigg\},
 \label{eq:app4-16}
 \end{eqnarray}
 \begin{eqnarray}
 {\cal A}^{a}({\overline{B}}_{s}^{0}{\to}{\pi}^{+}{\rho}^{-})
  &=&   \frac{G_{F}}{\sqrt{2}} f_{B_{s}} f_{\pi} f_{\rho} \bigg\{
        V_{ub}V_{us}^{\ast} \Big[ b_{1}({\pi}^{+},{\rho}^{-})
  \nonumber \\
  & & 
      + \Big( V_{ub}V_{us}^{\ast} + V_{cb}V_{cs}^{\ast} \Big)
        \Big[ b_{4}({\pi}^{+},{\rho}^{-})
      + b_{4}({\rho}^{-},{\pi}^{+})
  \nonumber \\
  & & 
      + b_{4}^{ew}({\pi}^{+},{\rho}^{-})
      - \frac{1}{2} b_{4}^{ew}({\rho}^{-},{\pi}^{+})
        \Big] \bigg\},
 \label{eq:app4-17}
 \end{eqnarray}
 \begin{eqnarray}
 {\cal A}^{a}({\overline{B}}_{s}^{0}{\to}{\pi}^{-}{\rho}^{+})
  &=&   \frac{G_{F}}{\sqrt{2}} f_{B_{s}} f_{\pi} f_{\rho} \bigg\{
        V_{ub}V_{us}^{\ast} \Big[ b_{1}({\rho}^{+},{\pi}^{-})
  \nonumber \\
  & & 
      + \Big( V_{ub}V_{us}^{\ast} + V_{cb}V_{cs}^{\ast} \Big)
        \Big[ b_{4}({\rho}^{+},{\pi}^{-})
      + b_{4}({\pi}^{-},{\rho}^{+})
  \nonumber \\
  & & 
      + b_{4}^{ew}({\rho}^{+},{\pi}^{-})
      - \frac{1}{2} b_{4}^{ew}({\pi}^{-},{\rho}^{+})
        \Big] \bigg\},
 \label{eq:app4-18}
 \end{eqnarray}
 \begin{eqnarray}
 {\cal A}^{a}({\overline{B}}_{s}^{0}{\to}{\pi}^{-}K^{{\ast}+})
  &=&   \frac{G_{F}}{\sqrt{2}} f_{B_{s}} f_{\pi} f_{K^{\ast}}
        \Big( V_{ub}V_{ud}^{\ast} + V_{cb}V_{cd}^{\ast} \Big)
        \bigg\{ b_{3}({\pi}^{-},K^{{\ast}+})
      - \frac{1}{2} b_{3}^{ew}({\pi}^{-},K^{{\ast}+}) \bigg\},
 \label{eq:app4-19}
 \end{eqnarray}

 \end{appendix}

 \begin{table}[ht]
 \caption{Wilson coefficients $C_{i}$ in the NDR scheme.
          Input parameters in numerical calculations are:
          ${\alpha}_{s}(m_{Z}) =0.117$,
          ${\alpha}_{em}(m_{W})=1/128$,
          $m_{W}=80.42 \,\text{GeV}$,
          $m_{Z}=91.188\,\text{GeV}$,
          $m_{t}=178.1 \,\text{GeV}$,
          $m_{b}=4.66  \,\text{GeV}$.}
 \label{tab1}
 \begin{ruledtabular}
 \begin{tabular}{crrrrrr}
            & \multicolumn{2}{c}{${\mu}=m_{b}/2 $}
            & \multicolumn{2}{c}{${\mu}=m_{b}   $}
            & \multicolumn{2}{c}{${\mu}=2m_{b}  $} \\
              \cline{2-3} \cline{4-5} \cline{6-7}
            & NLO & LO & NLO & LO & NLO & LO \\ \hline
  $C_{1} $  & 1.130 & 1.171 & 1.078 & 1.111 & 1.042 & 1.071 \\
  $C_{2} $  &-0.274 &-0.342 &-0.176 &-0.239 &-0.102 &-0.161 \\
  $C_{3} $  & 0.021 & 0.019 & 0.014 & 0.012 & 0.009 & 0.007 \\
  $C_{4} $  &-0.048 &-0.047 &-0.034 &-0.032 &-0.024 &-0.022 \\
  $C_{5} $  & 0.010 & 0.010 & 0.008 & 0.008 & 0.007 & 0.006 \\
  $C_{6} $  &-0.061 &-0.058 &-0.039 &-0.037 &-0.026 &-0.023 \\
  $C_{7}/{{\alpha}_{em}} $ &-0.005 &-0.105 &-0.011 &-0.097 & 0.035 &-0.081 \\
  $C_{8}/{{\alpha}_{em}} $ & 0.086 & 0.023 & 0.055 & 0.014 & 0.036 & 0.009 \\
  $C_{9}/{{\alpha}_{em}} $ &-1.419 &-0.091 &-1.341 &-0.087 &-1.277 &-0.075 \\
  $C_{10}/{{\alpha}_{em}}$ & 0.383 &-0.021 & 0.264 &-0.016 & 0.176 &-0.011 \\
  $C_{7{\gamma}}         $ &       &-0.342 &       &-0.306 &       &-0.276 \\
  $C_{8g}                $ &       &-0.160 &       &-0.146 &       &-0.133 \\
 \end{tabular}
 \end{ruledtabular}
 \end{table}

 \begin{table}[ht]
 \caption{The values of the Wolfenstein parameters, $A$, ${\lambda}$,
         ${\rho}$, and ${\eta}$}
 \label{tab2}
 \begin{ruledtabular}
 \begin{tabular}{ccccc}
   Refs.     & \cite{0012308} & \cite{0104062}
             & \cite{0207101} & \cite{pdg2002} \\ \hline
 ${\lambda}$ & 0.2237 ${\pm}$ 0.0033
             & 0.222  ${\pm}$ 0.004
             & 0.2210 ${\pm}$ 0.0020
             & 0.2236 ${\pm}$ 0.0031 \footnotemark[1]
     \footnotetext[1]{Determined from the measurements of
     ${\vert}V_{ud}{\vert}$ = 0.9734 ${\pm}$ 0.0008 and
     ${\vert}V_{us}{\vert}$ = 0.2196 ${\pm}$ 0.0026, and
     the error includes scale factor of 1.5}\\
 $A$         & 0.819  ${\pm}$ 0.040 \footnotemark[2]
     \footnotetext[2]{$A={\vert}V_{cb}{\vert}/{\lambda}$, and
     ${\vert}V_{cb}{\vert}$ =
     $(40.6{\pm}0.8){\times}10^{-3}$ \cite{0207101},
     $(41.2{\pm}2.0){\times}10^{-3}$ \cite{pdg2002}.}
             & 0.83   ${\pm}$ 0.07
             & 0.831  ${\pm}$ 0.022 \footnotemark[2]
             & 0.824  ${\pm}$ 0.046 \footnotemark[2] \\
 $\bar{\rho}$ \footnotemark[3]
    \footnotetext[3]{$\bar{\rho}={\rho}(1-{\lambda}^{2}/2)$}
             & 0.224  ${\pm}$ 0.038
             & 0.21   ${\pm}$ 0.12
             & 0.173  ${\pm}$ 0.046
             & 0.22   ${\pm}$ 0.10  \\
 $\bar{\eta}$ \footnotemark[4]
    \footnotetext[4]{$\bar{\eta}={\eta}(1-{\lambda}^{2}/2)$}
             & 0.317  ${\pm}$ 0.040
             & 0.38   ${\pm}$ 0.11
             & 0.357  ${\pm}$ 0.027
             & 0.35   ${\pm}$ 0.05  \\
 ${\gamma}$  & $(54.8{\pm}6.2)^{\circ}$
             & $(62{\pm}15)^{\circ}$
             & $(63.5{\pm}7.0)^{\circ}$
             & $(59{\pm}13)^{\circ}$
 \end{tabular}
 \end{ruledtabular}
 \end{table}

 \begin{table}[ht]
 \caption{Values of meson decay constants, form factors, and
 ${\eta}$-${\eta}^{\prime}$ mixing parameters.}
 \label{tab3}
 \begin{ruledtabular}
 \begin{tabular}{rll|rll|rll|rll}
 \multicolumn{3}{c|}{Form factor}     &
 \multicolumn{6}{c|}{Decay constants} &
 \multicolumn{3}{c}{${\eta}$-${\eta}^{\prime}$ mixing angles} \\ \hline
 $F_{0}^{B_{s}K}$ & 0.274            & \cite{prd48}   &
 $f_{\pi}$        & 131 MeV          & \cite{pdg2002} &
 $f_{K^{\ast}}$   & 214 MeV          & \cite{9804363} &
 ${\theta}$       & - $15.4^{\circ}$ & \cite{9802409} \\
 $F_{0}^{B_{s}{\eta}_{s\bar{s}}}$  & 0.335 & \cite{prd48} &
 $f_{K}$          & 160 MeV          & \cite{pdg2002} &
 $f_{\rho}$       & 210 MeV          & \cite{9804363} &
 ${\theta}_{0}$   & - $ 9.2^{\circ}$ & \cite{9802409} \\
 $F_{0}^{B_{s}{\eta}^{\prime}_{s\bar{s}}}$  & 0.282 & \cite{prd48} &
 $f_{0}$          & 1.17 $f_{\pi}$   & \cite{9802409} &
 $f_{\omega}$     & 195 MeV          & \cite{9804363} &
 ${\theta}_{8}$   & - $21.2^{\circ}$ & \cite{9802409} \\
 $A_{0}^{B_{s}K^{\ast}}$ & 0.236     & \cite{prd48}   &
 $f_{8}$          & 1.26 $f_{\pi}$   & \cite{9802409} &
 $f_{\phi}$       & 233 MeV          & \cite{9804363} \\
 $A_{0}^{B_{s}{\phi}}$   & 0.272     & \cite{prd48}   &
 $f_{B_{s}}$      & 236 MeV          & \cite{0108242} & & & &
 \end{tabular}
 \end{ruledtabular}
 \end{table}

 \begin{table}[ht]
 \caption{The CP-averaged branching ratios (in the unit of $10^{-6}$)
  of decays $B_{s}$ ${\to}$ $PP$ calculated with ${\mu}$ = $m_{b}$,
  and default values of parameters. The data in the column $2{\sim}4$
  are computed with $A=0.824$, ${\lambda}=0.2236$, ${\bar{\rho}}=0.22$,
  ${\bar{\eta}}=0.35$, ${\gamma}=59^{\circ}$,
  while the data in the column $5{\sim}7$ are computed with
  $A=0.82$, ${\lambda}=0.22$, ${\bar{\rho}}=0.086$,
  ${\bar{\eta}}=0.39$, ${\gamma}=78.8^{\circ}$.}
 \label{tab4}
 \begin{ruledtabular}
 \begin{tabular}{lcccccc}
     \multicolumn{1}{c}{decay}
   & NF & \multicolumn{2}{c}{QCDF}
   & NF & \multicolumn{2}{c}{QCDF}
   \\ \cline{2-2}\cline{3-4}\cline{5-5}\cline{6-7}
   \multicolumn{1}{c}{modes}
   & $BR$ & $BR^{f}$ & $BR^{f+a}$
   & $BR$ & $BR^{f}$ & $BR^{f+a}$ \\ \hline
   ${\overline{B}}^{0}_{s}{\to}K^{0}{\overline{K}}^{0}$
   & 8.724 & 12.06 & 18.81
   & 7.995 & 11.06 & 17.25 \\
   ${\overline{B}}^{0}_{s}{\to}K^{0}{\pi}^{0}$
   & 0.189 & 0.154 & 0.200
   & 0.247 & 0.210 & 0.277 \\
   ${\overline{B}}^{0}_{s}{\to}K^{0}{\eta}$
   & 0.108 & 0.061 & 0.071
   & 0.125 & 0.077 & 0.091 \\
   ${\overline{B}}^{0}_{s}{\to}K^{0}{\eta}^{\prime}$
   & 0.434 & 0.497 & 0.717
   & 0.396 & 0.522 & 0.777 \\
   ${\overline{B}}^{0}_{s}{\to}{\pi}^{0}{\pi}^{0}$
   & --- & --- & 0.011
   & --- & --- & 0.011 \\
   ${\overline{B}}^{0}_{s}{\to}{\pi}^{0}{\eta}$
   & 0.052 & 0.078 & 0.087
   & 0.058 & 0.066 & 0.071 \\
   ${\overline{B}}^{0}_{s}{\to}{\pi}^{0}{\eta}^{\prime}$
   & 0.055 & 0.059 & 0.056
   & 0.061 & 0.062 & 0.062 \\
   ${\overline{B}}^{0}_{s}{\to}{\eta}{\eta}$
   & 4.570 & 7.084 & 10.52
   & 4.087 & 6.535 & 9.720 \\
   ${\overline{B}}^{0}_{s}{\to}{\eta}{\eta}^{\prime}$
   & 9.190 & 10.60 & 16.44
   & 8.401 & 9.655 & 14.96 \\
   ${\overline{B}}^{0}_{s}{\to}{\eta}^{\prime}{\eta}^{\prime}$
   & 4.622 & 6.034 & 10.65
   & 4.324 & 5.583 & 9.846 \\
   ${\overline{B}}^{0}_{s}{\to}K^{+}{\pi}^{-}$
   & 9.653 & 10.23 & 10.44
   & 7.311 & 7.667 & 7.700 \\
   ${\overline{B}}^{0}_{s}{\to}K^{+}K^{-}$
   & 6.949 & 9.762 & 15.63
   & 7.807 & 10.65 & 16.58 \\
   ${\overline{B}}^{0}_{s}{\to}{\pi}^{+}{\pi}^{-}$
   & --- & --- & 0.022
   & --- & --- & 0.023
 \end{tabular}
 \end{ruledtabular}
 \end{table}

 \begin{table}[ht]
 \caption{The CP-averaged branching ratios (in unit of $10^{-6}$) of
  decays $B_{s}$ ${\to}$ $PV$ calculated with ${\mu}$ = $m_{b}$, and
  default values of parameters. The data in the column $2{\sim}4$ are
  computed with $A=0.824$, ${\lambda}=0.2236$, ${\bar{\rho}}=0.22$,
  ${\bar{\eta}}=0.35$, ${\gamma}=59^{\circ}$,
  while the data in the column $5{\sim}7$ are computed with
  $A=0.82$, ${\lambda}=0.22$, ${\bar{\rho}}=0.086$,
  ${\bar{\eta}}=0.39$, ${\gamma}=78.8^{\circ}$.}
 \label{tab5}
 \begin{ruledtabular}
 \begin{tabular}{lcccccc}
     \multicolumn{1}{c}{decay}
   & NF & \multicolumn{2}{c}{QCDF}
   & NF & \multicolumn{2}{c}{QCDF}
   \\ \cline{2-2}\cline{3-4}\cline{5-5}\cline{6-7}
   \multicolumn{1}{c}{modes}
   & $BR$ & $BR^{f}$ & $BR^{f+a}$
   & $BR$ & $BR^{f}$ & $BR^{f+a}$ \\ \hline
   ${\overline{B}}^{0}_{s}{\to}K^{0}{\overline{K}}^{{\ast}0}$
   & 2.163 & 2.812 & 4.985
   & 1.982 & 2.581 & 4.577 \\
   ${\overline{B}}^{0}_{s}{\to}{\overline{K}}^{0}K^{{\ast}0}$
   & 0.422 & 0.630 & 2.381
   & 0.387 & 0.577 & 2.182 \\
   ${\overline{B}}^{0}_{s}{\to}K^{+}K^{{\ast}-}$
   & 1.818 & 2.254 & 3.819
   & 2.598 & 3.119 & 4.992 \\
   ${\overline{B}}^{0}_{s}{\to}K^{-}K^{{\ast}+}$
   & 1.262 & 1.606 & 3.906
   & 0.863 & 1.103 & 2.864 \\
   ${\overline{B}}^{0}_{s}{\to}{\pi}^{+}{\rho}^{-}$
   & --- & --- & 0.002
   & --- & --- & 0.001 \\
   ${\overline{B}}^{0}_{s}{\to}{\pi}^{-}{\rho}^{+}$
   & --- & --- & 0.002
   & --- & --- & 0.001 \\
   ${\overline{B}}^{0}_{s}{\to}K^{0}{\rho}^{0}$
   & 0.368 & 0.367 & 0.376
   & 0.349 & 0.350 & 0.402 \\
   ${\overline{B}}^{0}_{s}{\to}K^{0}{\omega}$
   & 0.422 & 0.428 & 0.480
   & 0.292 & 0.314 & 0.341 \\
   ${\overline{B}}^{0}_{s}{\to}K^{0}{\phi}$
   & 0.029 & 0.057 & 0.154
   & 0.036 & 0.070 & 0.188 \\
   ${\overline{B}}^{0}_{s}{\to}{\pi}^{0}K^{{\ast}0}$
   & 0.137 & 0.068 & 0.102
   & 0.095 & 0.052 & 0.082 \\
   ${\overline{B}}^{0}_{s}{\to}{\pi}^{0}{\rho}^{0}$
   & --- & --- & 0.002
   & --- & --- & 0.001 \\
   ${\overline{B}}^{0}_{s}{\to}{\pi}^{0}{\omega}$
   & --- & --- & 0.001
   & --- & --- & 0.001 \\
   ${\overline{B}}^{0}_{s}{\to}{\pi}^{0}{\phi}$
   & 0.086 & 0.100 & ---
   & 0.095 & 0.098 & --- \\
   ${\overline{B}}^{0}_{s}{\to}{\eta}K^{{\ast}0}$
   & 0.121 & 0.120 & 0.230
   & 0.169 & 0.154 & 0.300 \\
   ${\overline{B}}^{0}_{s}{\to}{\eta}^{\prime}K^{{\ast}0}$
   & 0.062 & 0.028 & 0.038
   & 0.043 & 0.021 & 0.028 \\
   ${\overline{B}}^{0}_{s}{\to}{\eta}{\rho}^{0}$
   & 0.122 & 0.207 & 0.223
   & 0.136 & 0.166 & 0.174 \\
   ${\overline{B}}^{0}_{s}{\to}{\eta}^{\prime}{\rho}^{0}$
   & 0.128 & 0.128 & 0.125
   & 0.144 & 0.146 & 0.148 \\
   ${\overline{B}}^{0}_{s}{\to}{\eta}{\omega}$
   & 0.042 & 0.043 & 0.055
   & 0.052 & 0.031 & 0.039 \\
   ${\overline{B}}^{0}_{s}{\to}{\eta}^{\prime}{\omega}$
   & 0.044 & 0.012 & 0.016
   & 0.055 & 0.014 & 0.017 \\
   ${\overline{B}}^{0}_{s}{\to}{\eta}{\phi}$
   & 0.259 & 0.530 & 0.456
   & 0.216 & 0.483 & 0.417 \\
   ${\overline{B}}^{0}_{s}{\to}{\eta}^{\prime}{\phi}$
   & 0.007 & 0.266 & 0.367
   & 0.005 & 0.238 & 0.328 \\
   ${\overline{B}}^{0}_{s}{\to}K^{+}{\rho}^{-}$
   & 23.69 & 24.12 & 24.36
   & 19.02 & 19.29 & 19.18 \\
   ${\overline{B}}^{0}_{s}{\to}{\pi}^{-}K^{{\ast}+}$
   & 6.643 & 6.949 & 6.910
   & 5.709 & 6.008 & 6.161
 \end{tabular}
 \end{ruledtabular}
 \end{table}

 \begin{table}[ht]
 \caption{The CP-violating asymmetry parameters $a_{{\epsilon}^{\prime}}$
 and $a_{{\epsilon}+{\epsilon}^{\prime}}$ for $B_{s}$ ${\to}$ $PP$ decays
 (in the unit of percent) calculated with the QCDF approach, with ${\mu}$
 = $m_{b}$, and default values of various parameters. The data in the
 column $2{\sim}5$ are computed with $A=0.824$, ${\lambda}=0.2236$,
 ${\bar{\rho}}=0.22$,  ${\bar{\eta}}=0.35$, ${\gamma}=59^{\circ}$,
 while the data in the column $6{\sim}9$ are computed with $A=0.82$,
 ${\lambda}=0.22$, ${\bar{\rho}}=0.086$, ${\bar{\eta}}=0.39$,
 ${\gamma}=78.8^{\circ}$.}
 \label{tab6}
 \begin{ruledtabular}
 \begin{tabular}{ccccccccc}
     \multicolumn{1}{c}{modes}
   & $a_{{\epsilon}^{\prime}}^{f}$
   & $a_{{\epsilon}^{\prime}}^{f+a}$
   & $a_{{\epsilon}+{\epsilon}^{\prime}}^{f}$
   & $a_{{\epsilon}+{\epsilon}^{\prime}}^{f+a}$
   & $a_{{\epsilon}^{\prime}}^{f}$
   & $a_{{\epsilon}^{\prime}}^{f+a}$
   & $a_{{\epsilon}+{\epsilon}^{\prime}}^{f}$
   & $a_{{\epsilon}+{\epsilon}^{\prime}}^{f+a}$ \\ \hline
     $B_{s}{\to}K^{0}{\overline{K}}^{0}$
   & -0.90 & -0.72  & 3.38 & 3.43
   & -0.98 & -0.79  & 3.66 & 3.72 \\
     $B_{s}{\to}K^{0}_{S}{\pi}^{0}$
   & -64.72 & -59.92 & -61.67 & -74.32
   & -47.54 & -43.46 & -86.01 & -89.93 \\
     $B_{s}{\to}K^{0}_{S}{\eta}$
   & -85.03 & -82.76 & -37.31 & -48.06
   & -67.24 & -64.29 & -71.02 & -75.70 \\
     $B_{s}{\to}K^{0}_{S}{\eta}^{\prime}$
   & 54.84 & 47.19 & -32.54 & -42.34
   & 52.27 & 43.60 & -39.40 & -47.14 \\
     $B_{s}{\to}{\pi}^{0}{\pi}^{0}$
   & --- & 0 & --- & -27.71
   & --- & 0 & --- & -28.25 \\
     $B_{s}{\to}{\pi}^{0}{\eta}$
   & -16.27 & -15.71 & 24.07 & 35.66
   & -19.20 & -19.39 & 26.41 & 40.18 \\
     $B_{s}{\to}{\pi}^{0}{\eta}^{\prime}$
   & -22.84 & -22.75 & -31.39 & -44.93
   & -21.67 & -20.47 & -32.69 & -45.21 \\
     $B_{s}{\to}{\eta}{\eta}$
   & 1.08 & 0.86 & 1.97 & 1.57
   & 1.18 & 0.93 & 2.12 & 1.69 \\
     $B_{s}{\to}{\eta}{\eta}^{\prime}$
   & -0.69 & -0.54 & 5.14 & 5.49
   & -0.76 & -0.59 & 5.59 & 5.98 \\
     $B_{s}{\to}{\eta}^{\prime}{\eta}^{\prime}$
   & -3.25 & -2.47 & 1.22 & 1.40
   & -3.52 & -2.67 & 1.29 & 1.50 \\
     $B_{s}{\to}K^{\pm}K^{\mp}$
   & -6.89 & -5.33 & -40.89 & -33.57
   & -6.32 & -5.04 & -40.74 & -33.89 \\
     $B_{s}{\to}{\pi}^{\pm}{\pi}^{\mp}$
   & --- & 0 & --- & -27.71
   & --- & 0 & --- & -28.25
 \end{tabular}
 \end{ruledtabular}
 \end{table}

 \begin{table}[ht]
 \caption{The CP-violating asymmetry parameters $a_{{\epsilon}^{\prime}}$
 and $a_{{\epsilon}+{\epsilon}^{\prime}}$ for $B_{s}$ ${\to}$ $PV$ decays
 (in the unit of percent) calculated with the QCDF approach, with ${\mu}$
 = $m_{b}$, and default values of various parameters. The data in the
 column $2{\sim}5$ are computed with $A=0.824$, ${\lambda}=0.2236$,
 ${\bar{\rho}}=0.22$, ${\bar{\eta}}=0.35$, ${\gamma}=59^{\circ}$,
 while the data in the column $6{\sim}9$ are computed with $A=0.82$,
 ${\lambda}=0.22$, ${\bar{\rho}}=0.086$, ${\bar{\eta}}=0.39$,
 ${\gamma}=78.8^{\circ}$.}
 \label{tab7}
 \begin{ruledtabular}
 \begin{tabular}{ccccccccc}
     \multicolumn{1}{c}{modes}
   & $a_{{\epsilon}^{\prime}}^{f}$
   & $a_{{\epsilon}^{\prime}}^{f+a}$
   & $a_{{\epsilon}+{\epsilon}^{\prime}}^{f}$
   & $a_{{\epsilon}+{\epsilon}^{\prime}}^{f+a}$
   & $a_{{\epsilon}^{\prime}}^{f}$
   & $a_{{\epsilon}^{\prime}}^{f+a}$
   & $a_{{\epsilon}+{\epsilon}^{\prime}}^{f}$
   & $a_{{\epsilon}+{\epsilon}^{\prime}}^{f+a}$ \\ \hline
     $B_{s}{\to}K^{0}_{S}{\rho}^{0}$
   & -27.38 & -21.76 & 68.58 & 45.24
   & -28.72 & -20.39 &  4.69 &-24.52 \\
     $B_{s}{\to}K^{0}_{S}{\omega}$
   & 38.51 & 32.73 & 91.14 & 87.34
   & 52.52 & 46.18 & 79.67 & 88.70 \\
     $B_{s}{\to}K^{0}_{S}{\phi}$
   & 9.29 & 6.07 & -76.99 & -75.32
   & 7.58 & 4.99 & -74.30 & -72.78 \\
     $B_{s}{\to}{\pi}^{0}{\rho}^{0}$
   & --- & 0 & --- & 99.48
   & --- & 0 & --- & 92.23 \\
     $B_{s}{\to}{\pi}^{0}{\omega}$
   & --- & 0 & --- & 89.01
   & --- & 0 & --- & 39.27 \\
     $B_{s}{\to}{\pi}^{0}{\phi}$
   & -20.88 & --- & -13.09 & ---
   & -21.30 & --- & -14.58 & --- \\
     $B_{s}{\to}{\eta}{\rho}^{0}$
   & -15.06 & -11.51 & 38.43 & 45.56
   & -18.77 & -14.75 & 43.57 & 52.53 \\
     $B_{s}{\to}{\eta}^{\prime}{\rho}^{0}$
   & -25.73 & -30.00 & -52.01 & -60.37
   & -22.49 & -25.30 & -51.79 & -59.53 \\
     $B_{s}{\to}{\eta}{\omega}$
   & -18.83 & -12.24 & 68.45 & 74.11
   & -26.04 & -17.28 & 78.68 & 85.81 \\
     $B_{s}{\to}{\eta}^{\prime}{\omega}$
   & -71.72 & -59.77 & 12.78 & 38.52
   & -63.65 & -57.69 &-47.14 &-27.12 \\
     $B_{s}{\to}{\eta}{\phi}$
   & 6.56 & 7.14 & 4.68 & 4.13
   & 7.19 & 7.81 & 5.00 & 4.37 \\
     $B_{s}{\to}{\eta}^{\prime}{\phi}$
   & 10.95 & 8.88 &  9.91 &  9.71
   & 12.25 & 9.93 & 10.60 & 10.48
 \end{tabular}
 \end{ruledtabular}
 \end{table}

 \begin{table}[ht]
 \caption{The CP-violating asymmetry parameters ${\cal A}_{CP}$
 ($\%$) for $B_{s}$ ${\to}$ $PP$ decays calculated with the QCDF
 approach, with ${\mu}$ = $m_{b}$, and default values of various
 parameters. The data in the column $3{\sim}4$ are computed with
 $A=0.824$, ${\lambda}=0.2236$, ${\bar{\rho}}=0.22$,
 ${\bar{\eta}}=0.35$, ${\gamma}=59^{\circ}$,
 while the data in the column $5{\sim}6$ are computed with $A=0.82$,
 ${\lambda}=0.22$, ${\bar{\rho}}=0.086$, ${\bar{\eta}}=0.39$,
 ${\gamma}=78.8^{\circ}$.}
 \label{tab8}
 \begin{ruledtabular}
 \begin{tabular}{lccccc}
     \multicolumn{1}{c}{modes}
   & \multicolumn{1}{c}{case}
   & ${\cal A}_{CP}^{f}$ & ${\cal A}_{CP}^{f+a}$
   & ${\cal A}_{CP}^{f}$ & ${\cal A}_{CP}^{f+a}$ \\ \hline
     $B_{s}{\to}K^{0}{\overline{K}}^{0}$ & II
   & 0.17 & 0.17
   & 0.18 & 0.18 \\
     $B_{s}{\to}K^{0}_{S}{\pi}^{0}$ & II
   & -3.24 & -3.86
   & -4.41 & -4.59 \\
     $B_{s}{\to}K^{0}_{S}{\eta}$ & II
   & -2.07 & -2.60
   & -3.71 & -3.94 \\
     $B_{s}{\to}K^{0}_{S}{\eta}^{\prime}$ & II
   & -1.49 & -1.99
   & -1.83 & -2.24 \\
     $B_{s}{\to}{\pi}^{0}{\pi}^{0}$ & II
   & --- & -1.38
   & --- & -1.41 \\
     $B_{s}{\to}{\pi}^{0}{\eta}$ & II
   & 1.16 & 1.74
   & 1.27 & 1.96 \\
     $B_{s}{\to}{\pi}^{0}{\eta}^{\prime}$ & II
   & -1.62 & -2.30
   & -1.68 & -2.31 \\
     $B_{s}{\to}{\eta}{\eta}$ & II
   & 0.10 & 0.08
   & 0.11 & 0.09 \\
     $B_{s}{\to}{\eta}{\eta}^{\prime}$ & II
   & 0.25 & 0.27
   & 0.28 & 0.30 \\
     $B_{s}{\to}{\eta}^{\prime}{\eta}^{\prime}$ & II
   & 0.053 & 0.064
   & 0.056 & 0.068 \\
     $B_{s}{\to}K^{\pm}{\pi}^{\mp}$ & I
   & -4.42 & -5.09
   & -5.91 & -6.91 \\
     $B_{s}{\to}K^{\pm}K^{\mp}$ & II
   & -2.06 & -1.69
   & -2.05 & -1.70 \\
     $B_{s}{\to}{\pi}^{\pm}{\pi}^{\mp}$ & II
   & --- & -1.38
   & --- & -1.41
 \end{tabular}
 \end{ruledtabular}
 \end{table}

 \begin{table}[ht]
 \caption{The CP-violating asymmetry parameters ${\cal A}_{CP}$
 ($\%$) for $B_{s}$ ${\to}$ $PV$ decays calculated with the QCDF
 approach, with ${\mu}$ = $m_{b}$, and default values of various
 parameters. The data in the column $3{\sim}4$ are computed with
 $A=0.824$, ${\lambda}=0.2236$, ${\bar{\rho}}=0.22$,
 ${\bar{\eta}}=0.35$, ${\gamma}=59^{\circ}$,
 while the data in the column $5{\sim}6$ are computed with $A=0.82$,
 ${\lambda}=0.22$, ${\bar{\rho}}=0.086$, ${\bar{\eta}}=0.39$,
 ${\gamma}=78.8^{\circ}$.}
 \label{tab9}
 \begin{ruledtabular}
 \begin{tabular}{lccccc}
     \multicolumn{1}{c}{modes}
   & \multicolumn{1}{c}{case}
   & ${\cal A}_{CP}^{f}$ & ${\cal A}_{CP}^{f+a}$
   & ${\cal A}_{CP}^{f}$ & ${\cal A}_{CP}^{f+a}$ \\ \hline
     $B_{s}^{0}{\to}K^{0}_{S}{\overline{K}}^{{\ast}0}$ & III
   & 0.68 & 0.27
   & 0.74 & 0.29 \\
     $B_{s}^{0}{\to}K^{0}_{S}K^{{\ast}0}$ & III
   & -0.96 & -0.58
   & -1.05 & -0.63 \\
     $B_{s}^{0}{\to}K^{+}K^{{\ast}-}$ & III
   & 14.28 & -7.98
   & 12.02 & -8.50 \\
     $B_{s}^{0}{\to}K^{-}K^{{\ast}+}$ & III
   & -13.09 & 9.51
   & -13.04 & 8.71 \\
     $B_{s}^{0}{\to}{\pi}^{+}{\rho}^{-}$ & III
   & --- & 4.96
   & --- & 4.60 \\
     $B_{s}^{0}{\to}{\pi}^{-}{\rho}^{+}$ & III
   & --- & 4.96
   & --- & 4.60 \\
     $B_{s}{\to}K^{0}_{S}{\rho}^{0}$ & II
   & 3.35 &  2.20
   & 0.16 & -1.27 \\
     $B_{s}{\to}K^{0}_{S}{\omega}$ & II
   & 4.64 & 4.44
   & 4.10 & 4.54 \\
     $B_{s}{\to}K^{0}_{S}{\phi}$ & II
   & -3.82 & -3.74
   & -3.69 & -3.62 \\
     ${\overline{B}}^{0}_{s}{\to}{\pi}^{0}K^{{\ast}0}$ & I
   & -63.28 & -51.12
   & -81.61 & -63.76 \\
     $B_{s}{\to}{\pi}^{0}{\rho}^{0}$ & II
   & --- & 4.96
   & --- & 4.60 \\
     $B_{s}{\to}{\pi}^{0}{\omega}$ & II
   & --- & 4.44
   & --- & 1.96 \\
     $B_{s}{\to}{\pi}^{0}{\phi}$ & II
   & -0.70 & ---
   & -0.78 & --- \\
     ${\overline{B}}^{0}_{s}{\to}{\eta}K^{{\ast}0}$ & I
   & 49.72 & 30.43
   & 38.61 & 23.34 \\
     ${\overline{B}}_{s}{\to}{\eta}^{\prime}K^{{\ast}0}$ & I
   & -38.13 & -45.57
   & -50.43 & -62.38 \\
     $B_{s}{\to}{\eta}{\rho}^{0}$ & II
   & 1.88 & 2.24
   & 2.13 & 2.58 \\
     $B_{s}{\to}{\eta}^{\prime}{\rho}^{0}$ & II
   & -2.66 & -3.09
   & -2.64 & -3.03 \\
     $B_{s}{\to}{\eta}{\omega}$ & II
   & 3.37 & 3.67
   & 3.86 & 4.24 \\
     $B_{s}{\to}{\eta}^{\prime}{\omega}$ & II
   &  0.46 &  1.77
   & -2.51 & -1.50 \\
     $B_{s}{\to}{\eta}{\phi}$ & II
   & 0.25 & 0.22
   & 0.27 & 0.24 \\
     $B_{s}{\to}{\eta}^{\prime}{\phi}$ & II
   & 0.52 & 0.51
   & 0.56 & 0.55 \\
     $B_{s}{\to}K^{\pm}{\rho}^{\mp}$ & I
   & -2.09 &  0.79
   & -2.61 &  1.00 \\
     $B_{s}{\to}{\pi}^{\pm}K^{{\ast}{\mp}}$ & I
   & 0.06 & -5.10
   & 0.07 & -5.73
 \end{tabular}
 \end{ruledtabular}
 \end{table}

 \begin{table}[ht]
 \caption{Penguin-to-tree ratios and bound on ${\gamma}$, using the
         default values of various parameters, with the QCDF approach.
         The bound on ${\gamma}$ is from Eq. (\ref{eq:cosr}). The data
         in row a are computed with $A=0.824$, ${\lambda}=0.2236$,
         ${\bar{\rho}}=0.22$, ${\bar{\eta}}=0.35$, ${\gamma}=59^{\circ}$,
         while the data in row b are computed with $A=0.82$,
         ${\lambda}=0.22$, ${\bar{\rho}}=0.086$, ${\bar{\eta}}=0.39$,
         ${\gamma}=78.8^{\circ}$.}
 \label{tab10}
 \begin{ruledtabular}
 \begin{tabular}{cccccc}
    & ${\vert}P_{c}/T_{c}{\vert}$
    & ${\vert}P_{{\pi}{\pi}}/T_{{\pi}{\pi}}{\vert}$
    & ${\delta}^{\prime}$ & $R_{KK}$ & ${\gamma}$  \\ \hline
  a & 5.359 & 0.282 & $8.13^{\circ}$ & 0.831 & $<61.79^{\circ}$ \\ \hline
  b & 5.740 & 0.292 & $8.13^{\circ}$ & 0.961 & $<83.52^{\circ}$
 \end{tabular}
 \end{ruledtabular}
 \end{table}

 \end{document}